\documentclass[%
 reprint,
superscriptaddress,
nofootinbib,
 amsmath,amssymb,
 aps,
pra,
]{revtex4-2}

\usepackage{graphicx}
\usepackage{dcolumn}
\usepackage{bm}

\usepackage[colorlinks=true,
            linkcolor=magenta,
            citecolor=cyan,
            filecolor=magenta,
            urlcolor=magenta]{hyperref}
\usepackage{aas_macros}
\usepackage{multirow}

\begin{document}

\preprint{APS/123-QED}

\title{Expectations for the first supermassive black-hole binary resolved by PTAs\\ II: Milestones for binary characterization}

\author{Polina Petrov}
\email[Corresponding author: ]{polina.petrov@vanderbilt.edu}
\affiliation{Department of Physics and Astronomy, Vanderbilt University, 2301 Vanderbilt Place, Nashville, TN 37235, USA}

\author{Levi Schult}
\affiliation{Department of Physics and Astronomy, Vanderbilt University, 2301 Vanderbilt Place, Nashville, TN 37235, USA}

\author{Stephen~R.~Taylor}
\affiliation{Department of Physics and Astronomy, Vanderbilt University, 2301 Vanderbilt Place, Nashville, TN 37235, USA}

\author{Nihan Pol}
\affiliation{Department of Physics \& Astronomy, Texas Tech University, Lubbock, TX, 79409, USA}

\author{Nima Laal}
\affiliation{Department of Physics and Astronomy, Vanderbilt University, 2301 Vanderbilt Place, Nashville, TN 37235, USA}

\author{Maria Charisi}
\affiliation{Department of Physics and Astronomy, Washington State University, Pullman, WA 99163, USA}
\affiliation{Institute of Astrophysics, FORTH, GR-71110, Heraklion, Greece}

\author{Chung-Pei Ma}
\affiliation{Department of Astronomy, University of California, Berkeley, CA 94720, USA}
\affiliation{Department of Physics, University of California, Berkeley, CA 94720, USA}

\begin{abstract}
Following the recent evidence for a gravitational wave (GW) background found by pulsar timing array (PTA) experiments, the next major science milestone is resolving individual supermassive black hole binaries (SMBHBs). The detection of these systems could arise via searches using a power-based GW anisotropy model, i.e., a cross-correlation search for single sources, or a deterministic template model. In \citet{DelphiI}, we compared the efficacy of these models in constraining the GW signal from a single SMBHB using realistic, near-future PTA datasets, and found that the full-signal deterministic continuous wave (CW) search may achieve detection and characterization first. Here, we continue our analyses using only the CW model given its better performance, focusing now on characterization milestones. We examine the order in which CW parameters are constrained as PTA data are accumulated and the signal-to-noise ratio (S/N) grows. We also study how these parameter constraints vary across sources of different sky locations and GW frequencies. We find that the GW frequency and strain are generally constrained at the same time (or S/N), closely followed by the sky location, and later the chirp mass (if the source is highly evolving) and inclination angle. At fixed S/N, sources at higher frequencies generally achieve better precision on the GW frequency, chirp mass, and sky location. The time (and S/N) at which the signal becomes constrained is dependent on the sky location and frequency of the source, with the effects of pulsar terms and PTA geometry playing crucial roles in source detection and localization.

\end{abstract}

\maketitle

\section{\label{sec:Intro}Introduction}

Pulsar timing arrays (PTAs) \cite{1978SvA....22...36S,1979ApJ...234.1100D,1990ApJ...361..300F} leverage the exceptional timing precision of millisecond pulsars to search for gravitational waves (GWs) at nanohertz frequencies. Through decades-long monitoring of a collection of pulsars and the times-of-arrival (TOAs) of their radio pulses, nanohertz GW signals can be detected from the correlated deviations they induce in the pulse TOAs. PTA collaborations worldwide, including the North American Nanohertz Observatory for Gravitational Waves (NANOGrav), the European Pulsar Timing Array (EPTA), the Indian Pulsar Timing Array (InPTA), the Parkes Pulsar Timing Array (PPTA), the Chinese Pulsar Timing Array (CPTA), and the MeerKAT Pulsar Timing Array, have recently reported evidence for a stochastic gravitational wave background (GWB) at the 2-4$\sigma$ level \cite{NG15_gwb,EPTA_InPTA_gwb,PPTA_gwb,CPTA_gwb,MPTA_gwb}. In the near future, further evidence of the GWB's presence may be seen in the International PTA's third data release (IPTA-DR3), which will recombine and reanalyze a subset of the latest PTA datasets and is expected to be more sensitive than any individual dataset \cite{2024ApJ...966..105A, 2025arXiv250320949L, 2022MNRAS.510.4873A}.

The source of the GWB is predicted to be, and consistent with, that of a cosmic population of supermassive black hole binary (SMBHB) systems \cite{1995ApJ...446..543R,2003ApJ...583..616J,2004ApJ...611..623S,2019A&ARv..27....5B,NG15_astrointerp,EPTA_DR2_astrointerp}. These systems are expected to form as natural products of galaxy mergers \cite{BBR1980}, first becoming dual active galactic nuclei at kiloparsec separations, and later becoming gravitationally bound at scales of a few parsecs \cite{2023arXiv231016896D}. SMBHBs eventually shrink down to milliparsec separations, at which GWs dominate the binary orbital evolution and are detectable in the PTA band. Each SMBHB produces its own individual GW signal, called a continuous wave (CW) due to its minimal frequency evolution \cite{SVV2009}, while the incoherent superposition of all signals from the millions-strong SMBHB population produces a low-frequency GWB. Not every SMBHB system is discoverable as a CW signal, however, as the GWB acts as a source of noise above which a CW signal must be resolvable. CWs can therefore be disentangled from the background only if they come from particularly high-mass systems and/or systems which lie in the local Universe out to $z \sim 0.5$ \cite{Rosado15,Kelley18,BCK22_realgwbs,2025CQGra..42b5021C,Nicco2025}. Although CW searches conducted in PTA datasets have not yet turned up compelling evidence for such a system \cite{NG15_cw,EPTA_InPTA_DR2_cw, 2025arXiv250820007C,NG15_targetedCW}, a CW detection is nonetheless expected to follow a GWB detection \cite{Rosado15,Mingarelli17,Kelley18,BCK22_realgwbs}.

Individually-resolvable binaries should also induce point-source anisotropies in the GWB, appearing as “hotspots" of excess GW power on the sky \cite{2020PhRvD.102h4039T,Emiko24}. In \citet{DelphiI} (henceforth referred to as Paper I), we developed such a cross-correlation search for single sources and compared its efficacy to the full-signal CW search, in terms of the models' ability to detect and characterize a single SMBHB. For a suite of simulated PTA datasets segmented in time, we found that the CW search ultimately outperformed the anisotropy search. The CW model not only returned higher detection statistics (Savage-Dickey Bayes Factors and log-likelihood ratios), but also estimated the binary parameters more quickly, constraining the signal at earlier time slices. On top of this, the computational cost of the full CW signal search was comparable to that of the anisotropy search. Still, anisotropy searches will certainly have utility in cross-validating the presence of individual binaries in PTA datasets, and offering a first-cut approach to transdimensional multi-SMBHB searches that is less computationally demanding than deterministic multi-CW searches.

Having identified the full-signal CW search as the analysis method with the greatest efficacy, we now investigate binary characterization milestones over time slices from 5--22~years, or up to signal-to-noise ratios (S/N) of $\sim$~20. CW detection and parameter estimation in PTA data have been long-investigated topics; for example, \citet{SV2010} looked at measurement precision of CW parameters using the Fisher information matrix formalism, and \citet{CC2010} incorporated pulsar distances and phases into the CW signal model, studying how this ``pulsar term" portion of the signal impacts parameter measurements (see also \cite{Lee2011,Zhu2016}). More recently, \citet{Emiko25} conducted a SMBHB population study which included CW parameter precision as it relates to detection or non-detection of a CW signal. 

In this paper, we use the IPTA-inspired simulations from Paper I to assess the order in which CW parameters may be constrained following a realistic detection. We study how the parameter constraints evolve over time-sliced datasets, which incorporate not only a growing timespan but also the continued addition of pulsars into the array; consequently, these time slices contain a CW signal gradually increasing in S/N. We also study the dependence of sky localization, and parameter estimation more generally, on the source's parameters, and the role that pulsar terms and PTA geometry play in parameter estimation. Understanding how parameter measurements evolve from one dataset to the next will have important implications for future CW detections and multi-messenger campaigns. Evolving constraints can inform host galaxy identification efforts, including strategies to reduce and rank potential hosts within the localization volume as well as assessments of the overlap between GW and electromagnetic (EM) information in galaxy surveys \cite{G19,Petrov24,BardatiI,BardatiII,2025arXiv250421145H,Truant25}.

The sections that follow are organized in this manner: In \autoref{sec:Methods} we briefly review the relevant methods from Paper I (which are also used here), including the continuous gravitational wave model, our simulated PTA datasets, and GW signal recovery and parameter estimation. We present in \autoref{sec:Results} our results on the time-- and S/N--evolution of binary parameter constraints and the factors influencing these constraints. Finally, we discuss caveats and implications in \autoref{sec:Discuss} and conclude in \autoref{sec:Conclus}.

\section{\label{sec:Methods}Methods}

This section revisits the salient details of the continuous wave model, the realistic simulations, and the GW analysis methods used in Paper I, as well as in this work. For any outstanding details not mentioned here, we refer the reader to Paper I. Throughout the paper we use natural units $G = c = 1$.

\subsection{Continuous gravitational wave model}

Each pulsar's residuals, obtained by subtracting the pulsar's best-fit timing solution from the measured pulse times-of-arrival (TOAs), are modeled as
\begin{equation}
\label{eq:residuals}
    \textbf{r} = \vec{\delta t} - \textbf{M}\vec{\epsilon} - \textbf{F}\vec{a} - \vec{s},
\end{equation}
\noindent in which $\vec{\delta t}$ is a vector of the observed TOAs, \textbf{M} is the design matrix describing the linearized timing model, $\vec{\epsilon}$ are linear offsets from the timing model parameters, \textbf{F} is the Fourier design matrix, and $\vec{a}$ are the Fourier coefficients. The likelihood for a deterministic signal is then
\begin{equation}\label{eq:likelihood}
    \mathcal{L} = p(\vec{\delta t}|\vec{\nu}) = \frac{\mathrm{exp} \left( -\frac{1}{2}(\vec{\delta t} - \vec{s}) ^T \mathbf{C}^{-1} (\vec{\delta t} - s) \right)}{\sqrt{\mathrm{det}(2 \pi \mathbf{C})}}
\end{equation}
\noindent in which the vector $\vec{\nu}$ contains parameters describing both the deterministic CW signal and noise processes \cite{QuickCW}. The noise covariance matrix $\mathbf{C} = \mathrm{\mathbf{N}+ \mathbf{TBT}^T}$ includes the white noise covariance matrix $\mathrm{\mathbf{N}}$, the total design matrix $\mathbf{T} = [\mathbf{M} \; \mathbf{F}]$, and a matrix of priors $\mathbf{B}$ on the unbounded timing-model offsets and Fourier coefficients $\vec{a}$. Within the $\mathbf{B}$ matrix, the priors on the Fourier coefficients are governed by power-law hyperparameters on the intrinsic pulsar red noise and GWB \cite{2021arXiv210513270T}. These power-laws are of the form
\begin{equation}
P = \frac{A_{\mathrm{RN}}^2}{12\pi^2}\left(\frac{f}{f_{\mathrm{1yr}}}\right)^{-\gamma_{\mathrm{RN}}} \mathrm{yr}^3,
\end{equation}
\noindent where $P$ is the power spectral density, $A_{\mathrm{RN}}$ is the pulsar red noise amplitude, $f_{\mathrm{1yr}}$ is 1/(1yr) in Hz, and $\gamma_{\mathrm{RN}}$ is the power-law spectral index. Likewise the GWB power-law is
\begin{equation}
P = \frac{A_{\mathrm{GWB}}^2}{12\pi^2}\left(\frac{f}{f_{\mathrm{1yr}}}\right)^{-\gamma_{\mathrm{GWB}}} \mathrm{yr}^3,
\end{equation}
\noindent where the amplitude $A_{\mathrm{GWB}}$ and the spectral index $\gamma_{\mathrm{GWB}}$ are common to all pulsars. Since we use the software \texttt{QuickCW} \cite{QuickCW} (described in more detail later in this section), the GWB is modeled without Hellings-Downs (HD) spatial correlations \cite{1983ApJ...265L..39H}, meaning that $\mathbf{C}$ is block-diagonal with respect to the pulsars in the array.

The vector $\vec{s}$ in \autoref{eq:residuals} is the deterministic timing deviation signal produced by an individual binary. This signal can be written as
\begin{equation} \label{eq:s_vector}
    \vec{s}(t, \hat{\Omega}) = F^+(\hat{\Omega})\Delta s_+(t) + F^{\times}(\hat{\Omega})\Delta s_{\times}(t),
\end{equation}
\noindent where $+$ (``plus") and $\times$ (``cross") are the polarization modes, and the antenna pattern functions $F^{+,\times}$ describe the pulsars' response to the GW source. The antenna patterns are a function of $\hat{\Omega}$, the unit vector pointing in the direction of GW propagation, from the GW source to the Solar System Barycenter:
\begin{equation} 
    \hat{\Omega} = -(\sin\theta\cos\phi)\hat{x} - (\sin\theta\sin\phi)\hat{y} - (\cos\theta)\hat{z}.
\end{equation}

\noindent This unit vector is in turn dependent on the sky location of the binary in spherical polar coordinates ($\theta$, $\phi$), and the right ascension $\alpha$ and declination $\delta$ can be related to these coordinates through ($\theta$, $\phi$) $=$ ($\pi/2 - \delta$, $\alpha$).

The terms $\Delta s_{+,\times}$ in \autoref{eq:s_vector} encapsulate the total effect on the timing residuals, which is the difference between the signal induced at the Earth (the “Earth term”) and at a given pulsar (the “pulsar term”). The full signal is thus
\begin{equation}
    \Delta s_{\{+,\times\}}(t) = s_{\{+,\times\}}(t_p) - s_{\{+,\times\}}(t_e).
\end{equation}

\noindent The terms $t_e$ and $t_p$ are the times at which the GW passes the Earth and Pulsar respectively, and are related to each other by
\begin{equation} \label{eq:t_p}
t_p = t_e - L_{p}(1+\hat{\Omega}\cdot\hat{p}),
\end{equation}
\noindent where $L_p$ is the distance to the pulsar, and $\hat{p}$ is a vector pointing from the Earth to the pulsar.

Assuming the binaries have circular orbits,  $s_{+,\times}$ are defined, at zeroth post-Newtonian order, as
\begin{multline}
\label{eqn:circcwresid_splus}
    s_+(t) = \frac{\mathcal{M}^{5/3}}{d_L \omega(t)^{1/3}} [\sin[2(\Phi(t)](1+\cos^2\iota)\cos2\psi \\
    + 2\cos[2\Phi(t)]\cos\iota\sin(2\psi)],
\end{multline}
\begin{multline}
\label{eqn:circcwresid_scross}
    s_{\times}(t) = \frac{\mathcal{M}^{5/3}}{d_L \omega(t)^{1/3}} [-\sin[2\Phi(t)]](1+\cos^2\iota)\sin(2\psi) \\
    + 2\cos[2\Phi(t)]\cos\iota \cos2\psi],
\end{multline}

\noindent where $d_L$ is the luminosity distance to the source, $\mathcal{M} \equiv (m_1m_2)^{3/5} / (m_1 + m_2)^{1/5}$ is the chirp mass, composed of the two black hole masses $m_1$ and $m_2$, and $\omega(t)$ is the Earth-term orbital angular frequency. Here $\mathcal{M}$ and $\omega$ are the observer-frame quantities and can be related to the rest-frame quantities via $\mathcal{M}_r = \mathcal{M}/(1 + z)$ and $\omega_r = \omega(1 + z)$, with $z$ being the source's redshift. Given that PTAs are currently only sensitive to individually-resolvable SMBHBs in the local universe, we assume $1 + z \simeq 1$. The binary inclination angle $\iota$ is the angle between the line of sight and the binary's orbital angular momentum, $\psi$ is the GW polarization angle, and $\Phi(t)$ is the orbital phase of the binary. 

The orbital angular frequency and phase at time $t$ are given by
\begin{equation}
\label{eq:omega_t}
    \omega(t) = \omega_0\left(1 - \frac{256}{5} \mathcal{M}^{5/3}\omega_0^{8/3}t\right)^{-3/8},
\end{equation}
\begin{equation}
\label{eq:Phi_t}
    \Phi(t) = \Phi_0 + \frac{1}{32}\mathcal{M}^{-5/3}[\omega_0^{-5/3} - \omega(t)^{-5/3}],
\end{equation}
\noindent in which $\omega_0$ is the initial Earth-term orbital angular frequency, related to the GW frequency by $\omega_0 = \omega(t_0) = \pi f_{\mathrm{GW}}$, and $\Phi_0$ is the initial orbital phase. The binary loses energy through GW emission, slowly causing its orbital frequency to evolve over time as \cite{PM1963,P1964}
\begin{equation} \label{eq:domega_dt}
\frac{\mathrm{d}\omega}{\mathrm{d}t} = \frac{96}{5}\mathcal{M}^{5/3}\omega(t)^{11/3}.
\end{equation}

\noindent Because PTA experiments only stretch over decade-scale timespans, it is reasonable to assume that there is minimal binary evolution over the length of our simulated datasets. We therefore make use of the approximations $\omega(t)\approx\omega_0$ and $\Phi(t)\approx\Phi_0+\omega_0t$ rather than using the full orbital evolution expressions given by \autoref{eq:omega_t} and \autoref{eq:Phi_t}. Although the GW signal is assumed to be monochromatic over $\sim$decades, this assumption cannot be made with regards to the light-travel time between the Earth and any pulsar in the array. The average pulsar distance is $\sim$ 1 kpc -- a light-travel time on the order of $10^3$ years -- which creates significant evolution between the Earth and pulsar terms. The Earth term parameters can be evolved backwards to obtain the orbital frequency and phase of the pulsar term via
\begin{equation}\label{eq:w_p}
    \omega_p(t_p) = \omega_0\left(1+\frac{256}{5}\mathcal{M}^{5/3}\omega_0^{8/3}L_p(1+\hat{\Omega}\cdot\hat{p})\right)^{-3/8},
\end{equation}
\begin{equation}\label{eq:phi_p}
    \Phi_p(t_p) = \Phi_0 + \frac{1}{32}\mathcal{M}^{-5/3}\left(\omega_0^{-5/3}-\omega_p(t_p)^{-5/3}\right).
\end{equation}

Finally, the chirp mass $\mathcal{M}$, distance $d_L$, and GW frequency $f_{\rm{GW}}$ are related to the overall strain amplitude $h_0$ through
\begin{equation} \label{eq:h}
    h_0 = \frac{2\mathcal{M}^{5/3}(\pi f_\mathrm{GW})^{2/3}}{d_L}
\end{equation}
\noindent which in turn is related to the polarization amplitudes $h_+$ and $h_{\times}$ within the total response $h(t)=F_+h_+ + F_{\times}h_{\times}$:
\begin{equation}\label{eq:h_+}
    h_+=\frac{h_0}{2}(1+\cos^2\iota)\cos\Phi(t), 
\end{equation}
\begin{equation}\label{eq:h_x}
    h_{\times}=h_0\cos\iota\sin\Phi(t).
\end{equation}

While the circular CW signal can be modeled deterministically with the nine global parameters $\{\theta, \phi, d_L, \mathcal{M}, f_{\rm{GW}}, \iota, \psi, \Phi_0, h_0\}$, the degeneracy between $d_L$, $\mathcal{M}$, $f_{\rm{GW}}$, and $h_0$ leads to a full description of the signal that needs only eight global parameters. The luminosity distance $d_L$ is typically excluded in CW searches in real PTA datasets; hence, we choose not to search over it in this work. The CW signal also requires 2$N_{\rm{pulsar}}$ parameters, the pulsar's distance $L_p$ and the binary orbital phase at the time when the GW passes the pulsar $\Phi_p$. Despite being a derived parameter, $\Phi_p$ is treated as a free parameter in our analyses, as this allows for more efficient MCMC exploration of the likelihood \cite{2013CQGra..30v4004E}.

In total, our model includes (2 GWB parameters) + (2$N_{\rm{pulsar}}$ RN parameters) + (8 + 2$N_{\rm{pulsar}}$ CW parameters) = 10 + 4$N_{\rm{pulsar}}$ parameters.

\subsection{Simulated datasets}

\begin{figure}[!t]
    \centering
    \includegraphics[width=0.5\textwidth]{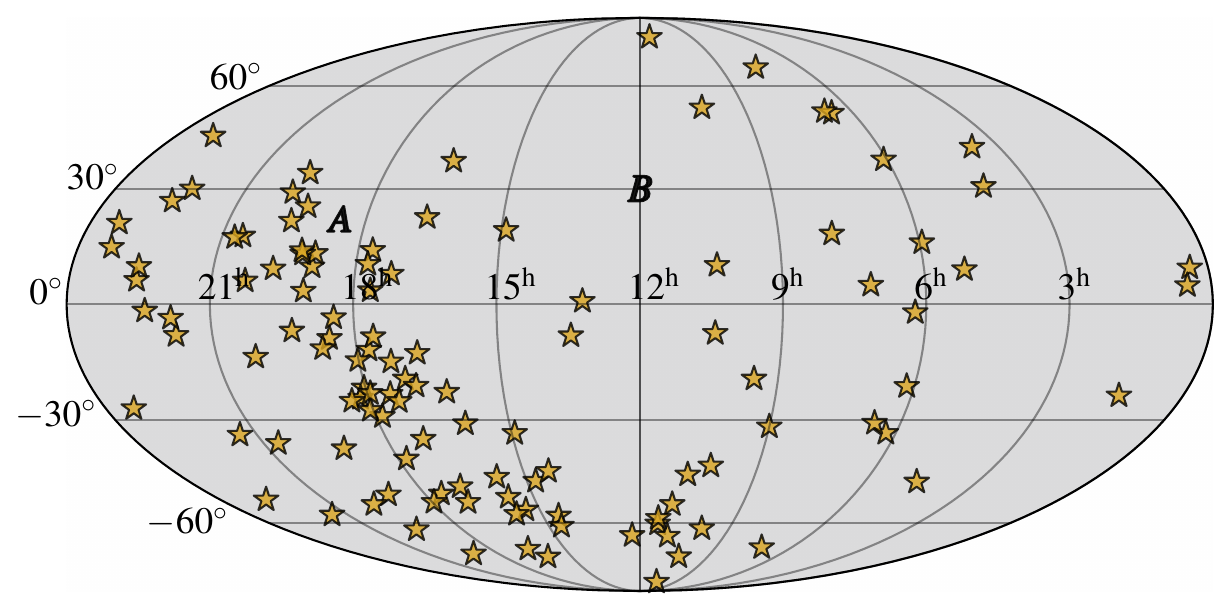}
    \caption{Sky map of our simulated PTA configuration (yellow stars) and the two sky locations A and B where we inject a CW signal.}
    \label{fig:skymap}
\end{figure}

We simulate a 116-pulsar array, shown by the yellow stars in \autoref{fig:skymap}, roughly emulating features of the near-future IPTA-DR3 dataset. The PTA configuration is based on realistic simulations presented in \citet{astro4cast} and extended to a 22-year baseline in \citet{Petrov24} (hereafter P24). We make use of the same simulated datasets and the same range of time slices and single-source S/N as those in Paper I; for a full description of the simulations, see details therein. Below we present a summary of the simulation procedure from these studies.

The simulated array includes 45 pulsars from the NANOGrav 12.5-year dataset \cite{2021ApJS..252....4A}; we use these pulsars' measured timing model parameters, TOAs, and TOA uncertainties, and additionally extend these observations to a 20-year baseline using statistics from each pulsar's final year of observations in the 12.5-year dataset. As for the other 71 pulsars, 23 pulsars come from the NANOGrav 15-year dataset \cite{NG15_dataset} (resulting in a total of 68 NANOGrav pulsars in this array), 14 pulsars are from PPTA DR3 \cite{2023arXiv230616230Z}, 3 pulsars are from EPTA$+$InPTA DR2new$+$ \cite{2023arXiv230616224A}, and 31 pulsars are from MPTA DR1 \cite{2023MNRAS.519.3976M}. Each of these 71 pulsars is positioned in its true location on the sky, but its timing model parameters are adopted from a randomized template pulsar among the pulsars J0931$-$1902, J1453$+$1902, J1832$-$0836, and J1911$+$1347. Observations are taken over the full baselines reported in each pulsar's respective dataset paper and then extended out to a 20-year NANOGrav baseline. Although the array's timing baseline reaches 20 years for the NANOGrav subset of pulsars, the baseline over all 116 pulsars is $\sim$22 years. The 71 pulsars' TOA uncertainties are taken to be the whitened RMS values from their respective dataset papers, and observations are taken at a two-week cadence. For choices made regarding pulsars timed by multiple PTAs, template pulsars from which to borrow timing model parameters, and other PTA configuration details, see P24 \cite{Petrov24} and Paper I.

We inject a CW signal into two sky locations, marked by letters A and B in \autoref{fig:skymap}. Sky location A lies in a region of the sky surrounded by many pulsars, where PTAs are generally more sensitive \cite{NG15_cw,Petrov24}, at coordinates $(\theta, \phi) = (1.19, 4.87)$ or $(\alpha,\delta) =$ (18h36m, $+22^{\circ}$). In contrast, sky location B lies in a region where pulsars are sparse and thus has lower sensitivity, at coordinates $(\theta, \phi) = (1.05, 3.14)$ or $(\alpha, \delta) =$ (12h, $+30^{\circ}$). We inject signals of two different GW frequencies, $f_{\rm{GW}} = 5$ nHz and $f_{\rm{GW}} = 20$ nHz, into each sky location. Throughout the paper, we refer to these injections as A5, A20, B5, and B20. Each injection has fixed parameters $\mathcal{M} = 10^9$~$\rm{M}_{\odot}$, $\iota = 0$ (face-on), $\psi=\pi/4$, and $\Phi_0=\pi/4$. The luminosity distance of each source is chosen to produce a signal with optimal S/N $\sim$ 20. The optimal S/N \cite{1998PhRvD..58f3001J,TDM2019_bhfg} is defined as a noise-weighted inner product
\begin{equation} \label{eq:snr_opt}
\mathrm{(S/N)_{opt}} = \sqrt{(\vec{s} | \vec{s})} = \sqrt{\vec{s}^T\mathbf{C}^{-1}\vec{s}}
\end{equation}
\noindent where $\vec{s}$ is the concatenated CW signal vector over all pulsars, and $\mathbf{C}$ is the PTA noise covariance matrix, which contains all white noise and intrinsic pulsar red noise terms as well as the spatially-correlated GWB. Our (S/N)$_{\rm opt}$ calculations include HD correlations; therefore, we utilize the full pulsar cross-correlation structure within $\mathbf{C}$, and $\vec{s}$ must be the total CW signal over all pulsars. However, when modeling the GWB as a common uncorrelated red noise (CURN) process, $\mathbf{C}$ is block-diagonal, and the total (S/N)$_{\rm opt}^2$ is simply a sum over the individual pulsars' (S/N)$_{\rm opt,p}^2$.

The white noise, quantified by the EFAC parameter in PTA literature, is set to unity. This value asserts that the reported TOA uncertainty is accurate and comes solely from template matching errors. The intrinsic red noise is injected using the parameters reported in each pulsar's respective dataset paper. The GWB is included in our simulations as a common red noise process with HD spatial correlations, characterized by $\gamma_{\rm{GWB}} = 13/3$ and $A_{\rm{GWB}} = 2.4 \times 10^{-15}$ as measured in \citet{NG15_gwb}. For each of the four CW injections (A5, A20, B5, B20), we create 10 different noise realizations, with white noise, pulsar red noise, and the GWB time-series all varying between realizations. 

Because we include these noise processes in our simulations, we quote the log-likelihood S/N \cite{2012ApJ...756..175E,TDM2019_bhfg}
\begin{equation}
\mathrm{(S/N)}_{\Lambda} = \sqrt{2(\delta\vec{t} | \vec{s}) - (\vec{s} | \vec{s})} = \sqrt{2\ln\Lambda},
\end{equation}
\noindent which has generally been established as a better proxy for CW detection compared to other S/N statistics such as the optimal S/N defined above \cite{2007PhRvD..75b3004P,Emiko25}. The log-likelihood S/N is named due to its relation to log-likelihood ratio 
\begin{equation}
\ln\Lambda = \ln\frac{\mathcal{L}(\delta\vec{t}|\vec{s})}{\mathcal{L}(\delta\vec{t}|0)},
\end{equation}

\noindent which compares the hypothesis of a signal being present in the residuals (in our case, we are interested in the CW signal $\vec{s}$), versus the hypothesis in which no signal is present (in our case, only the noise processes described earlier, including the GWB). We calculate (S/N)$_\Lambda$ from $\ln\Lambda = \ln\mathcal{L}_{\rm{CW+GWB}} - \ln\mathcal{L}_{\rm{GWB}}$ using the true, injected parameters.

Finally, as was done in Paper I, we segment our dataset into 11 different time slices. The earliest time slice has a 5-year baseline, followed by time slices with 7-, 10-, 11-, 12-, 13-, 14-, 15-, 17-, and 20-year baselines. The last time slice is then the full, original 22-year dataset. We note that all pulsar properties are kept constant across time slices, including their noise parameters and distances. The only change between time slices is therefore the pulsars' baselines. By injecting a CW signal at 2 sky locations and 2 GW frequencies, injecting 10 different noise realizations for each CW signal, and sectioning off each dataset into 11 time slices, we end up with a total of 440 simulated datasets.

\subsection{Signal recovery}\label{sec:Analyses}

We recover the CW signal and estimate the binary parameters with the Markov Chain Monte Carlo (MCMC) sampler \texttt{QuickCW} \cite{QuickCW}, extended from the PTA analysis software \texttt{ENTERPRISE} \cite{2019ascl.soft12015E} and specifically tailored to GW searches for individually-resolvable SMBHBs. The signal model and priors are assembled with \texttt{ENTERPRISE} and \texttt{enterprise\_extensions} \cite{e_e}, while \texttt{QuickCW} performs a custom likelihood calculation. This customization splits nuisance projection parameters ($\iota$, $\Phi_0$, $\psi$, $h_0$, $\Phi_p$) from those that influence GW signal morphology ($\theta$, $\phi$, $f_{\rm{GW}}$, $\mathcal{M}$, $L_p$, $A_{\rm{RN},p}$, $\gamma_{\rm{RN},p}$), enabling a drastic increase in computational speed. For further details, see \citet{QuickCW}. We also make use of parallel tempering for more efficient sampling of the complicated likelihood surface.

We fix the white noise term EFAC=1. The intrinsic pulsar RN, GWB, and CW parameters that we sample over are listed in \autoref{tab:priortable}, along with their priors. The RN and GWB parameters are limited to frequency bins $f_i = i/T$, where $T$ is the timespan of a given sliced dataset, and the frequencies $f_i$ include $f_1$,~...,~$f_{N_{\rm{yr}}}$. (For example, the 5-year slice is analyzed over frequency bins $f_1=1/\rm{(5 yr)}, ..., f_5=5/\rm{(5 yr)}$.) Aside from this difference between time slices, all analyses are conducted in the same manner.

\begin{table}
\centering
\caption{\label{tab:priortable} Parameter priors used in all analyses.}
\begin{tabular}{p{3cm} p{3cm}}
\hline
Parameter & Prior \\
\hline
\hline
$A_{\rm{RN},p}$ & Uniform($-$18, $-$11) \\
$\gamma_{\rm{RN},p}$ & Uniform(0, 7) \\
$A_{\rm{GWB}}$ & Uniform($-$18, $-$11) \\
$\gamma_{\rm{GWB}}$ & Uniform(0, 7) \\
$\cos\theta$ & Uniform($-$1, 1) \\
$\phi$ & Uniform(0, 2$\pi$) \\
$\log_{10}(f_{\rm{GW}}/\rm{Hz})$ & Uniform($-$9, $-$7.5) \\
$\log_{10}h_0$ & Uniform($-$18, $-$11) \\
$\log_{10}(\mathcal{M}/M_{\odot})$ & Uniform(7, 10) \\
$\cos\iota$ & Uniform($-$1, 1) \\
$\psi$ & Uniform(0, $\pi$) \\
$\Phi_0$ & Uniform(0, 2$\pi$) \\
$L_p$ & Normal($L_p$, $\sigma_{L_p}$) \\
$\Phi_p$ & Uniform(0, 2$\pi$) \\
\hline
\end{tabular}
\end{table}

Importantly, \texttt{QuickCW} does not currently support a GWB model with HD spatial correlations, but rather models the GWB as a CURN process. In this work, we ultimately chose not to employ the full CW+HD model, as it is computationally impractical. An extended discussion on this topic is included in Section V of Paper I, and we briefly review the major details in Section \ref{sec:D2} of this paper.

\section{\label{sec:Results}Results}
\subsection{\label{sec:Res1}Evolution of binary parameter constraints}

\begin{figure}[!ht]
     \centering
     \includegraphics[width=0.48\textwidth]{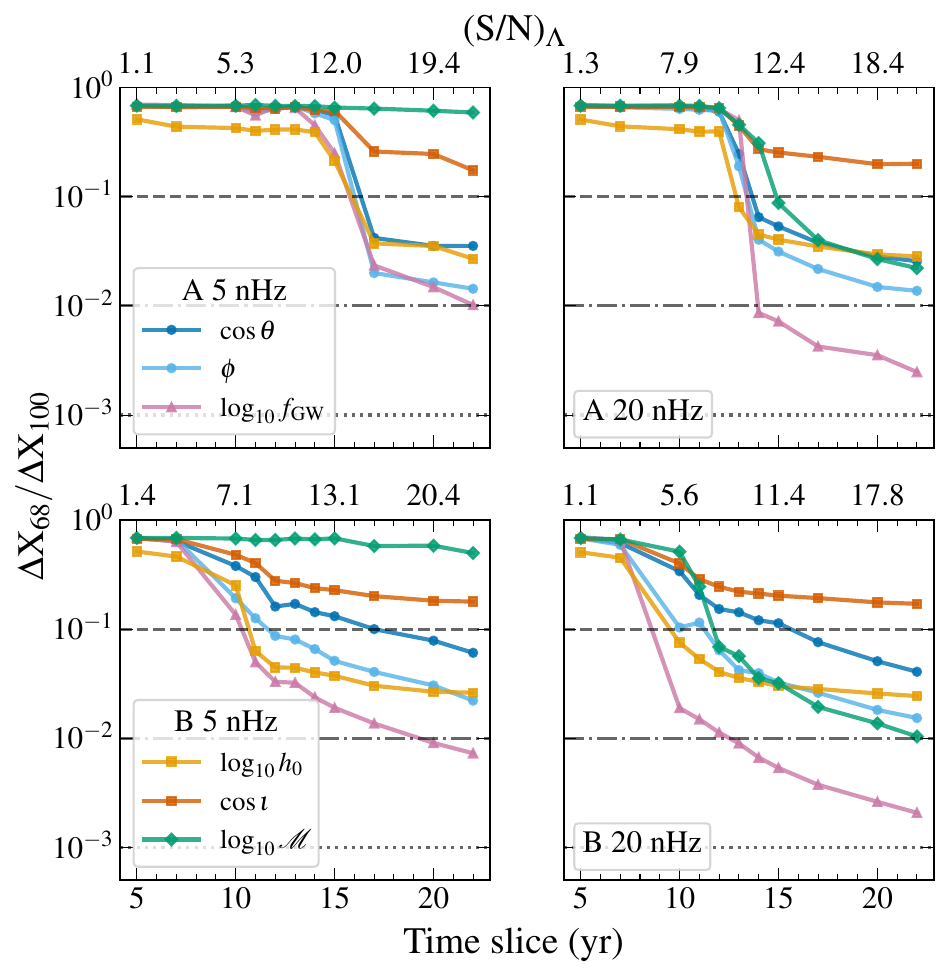}
     \caption{Parameter constraints a function of time slice and (S/N)$_\Lambda$ for all four simulation sets. For each simulation set, the constraints and (S/N)$_\Lambda$ shown are the median values across all noise realizations. The constraints are plotted as $\Delta X_{68}/\Delta X_{100}$, where $\Delta X_{68}$ is the 68\% credible interval of a given parameter $X$, and $\Delta X_{100}$ is the parameter's prior range. The sky location parameters $\cos\theta$ and $\phi$ are shown with dark and light blue circles, respectively. The GW frequency $\log_{10}f_{\rm{GW}}$ in shown with pink triangles, GW strain amplitude $\log_{10}h_0$ with gold squares, binary inclination angle $\cos\iota$ with orange squares, and chirp mass $\log_{10}\mathcal{M}$ with teal diamonds. The horizontal gray dashed, dash-dotted, and dotted lines indicate where the parameter is constrained to 10\%, 1\%, and 0.1\% of its prior, respectively.}
     \label{fig:qcw_par_order}
\end{figure}

We run the full Bayesian \texttt{QuickCW} analysis and estimate the binary parameters for each time-sliced dataset, tracking the parameter precision as pulsar timing data accumulate. In \autoref{fig:qcw_par_order}, we show the median parameter precision as a function of time slice (bottom x-axis) as well as the median (S/N)$_\Lambda$ (top x-axis). The top row of panels corresponds to Simulation A, while the bottom row corresponds to Simulation B, and the left and right columns correspond to the 5 nHz and 20 nHz injections, respectively. The quantity $\Delta X_{68}/\Delta X_{100}$ refers to the 68\% credible interval of a given parameter $X$ divided by that parameter's prior range; we plot this quantity for six parameters, including the polar angle $\cos\theta$ (dark blue circles), azimuthal angle $\phi$ (light blue circles), GW frequency $\log_{10}f_{\rm{GW}}$ (pink triangles), GW strain amplitude $\log_{10}h_0$ (gold squares), binary inclination angle $\cos\iota$ (orange squares), and chirp mass $\log_{10}\mathcal{M}$ (teal diamonds). 

\textit{Frequency and Strain} --- The first parameter to depart from the full range of its prior is generally the GW frequency $\log_{10}f_{\rm{GW}}$, followed closely by the strain amplitude $\log_{10}h_0$. This behavior is unsurprising, given that the frequency and amplitude influence the CW signal morphology. The two parameters usually start to become constrained at the same time slice or (S/N)$_\Lambda$, and their measurement precision improves monotonically with increasing slices. However, $\log_{10}f_{\rm{GW}}$ is measured more precisely from the outset and continues to achieve higher precision as data are accumulated. Its precision eventually reaches $\Delta X_{68}/\Delta X_{100} < 1\%$, whereas the precision on $\log_{10}h_0$ plateaus between $1-10\%$. In each injection, $\log_{10}f_{\rm{GW}}$ is the best-constrained parameter over the entire timespan of the dataset and is generally the only parameter to achieve $\Delta X_{68}/\Delta X_{100} < 1\%$.

Both parameters are expected to scale as (S/N)$^{-1}$ \cite{SV2010,Taylor2016}. The GW frequency roughly follows this behavior, while the strain appears to have a marginally shallower scaling. This slight deviation is likely due to our choice of face-on inclination angle for all CW injections; we discuss this point in more detail later in this section.

\textit{Sky Location} --- The sky location parameters $\cos\theta$ and $\phi$ generally become constrained simultaneously, and at the same or similar time slice as the frequency and strain. For both parameters, $\Delta X_{68}/\Delta X_{100}$ steadily decreases across time slices, arriving between $\sim 1-10\%$ by the final slice. Compared to the strain amplitude, the precision on $\phi$ is better (lower $\Delta X_{68}/\Delta X_{100}$) in each injection, while the precision on $\cos\theta$ is similar or slightly worse (higher $\Delta X_{68}/\Delta X_{100}$).

The azimuthal angle $\phi$ outperforms the polar angle $\cos\theta$ slightly, from the initial drop in measurement precision and beyond, as new data are added. This is in part due to the normalization by the prior $\Delta X_{100}$. The prior on $\phi \in [0,2\pi]$ results in $\Delta \phi_{100} = 2\pi$, while for $\cos\theta \in [-1,1]$ this is $\Delta \cos\theta_{100} = 2$. When comparing only $\Delta X_{68}$, i.e., when we do not normalize by the parameters' respective prior ranges, the precision is comparable. 

The difference in precision may also come from the sky coverage of the array. Source A sits at $\phi=4.87$ ($\alpha=$ 18h36m) and Source B sits at $\phi=3.14$ ($\alpha=$ 12h), both of whose azimuthal coordinates have noticeable gaps in $\theta$ as seen in \autoref{fig:skymap}. For location A this gap is roughly $\cos\theta > 0.5$ ($\delta > +30^{\circ}$), while for location B this gap is even larger, about $-0.82 > \cos\theta > 0.82$ ($-55^{\circ} > \delta > +55^{\circ}$). Accordingly, in \autoref{fig:qcw_par_order} the difference in $\Delta X_{68}/\Delta X_{100}$ between $\cos\theta$ and $\phi$ is somewhat more pronounced for B5 and B20, owing to location B's worse coverage of polar angle $\theta$ at $\phi=3.14$. On the contrary, taking a sliver in declination at sky location A ($\theta$=1.19, or $\delta=+22^{\circ}$) and sky location B ($\theta$=1.05, or $\delta=+30^{\circ}$) shows much more uniform coverage in azimuthal angle $\phi$. For a visual representation, see \autoref{fig:Acov} and \autoref{fig:Bcov} in the Appendix. In essence, both sources sit at Dec.\ coordinates where the pulsars are decently distributed in R.A., but sit at R.A. coordinates in which the pulsars are poorly distributed in Dec. The different distributions of pulsars on the sky could therefore result in better constraints on $\phi$ than $\cos\theta$.

Although here we investigate only the parameters that are directly sampled in the MCMC analysis, the uncertainties on the sky location parameters can be translated into a single localization error-box on the sky $\Delta\Omega$. Comparing $\Delta\Omega_{68}/\Delta\Omega_{100}$ against the other parameters shown in \autoref{fig:qcw_par_order}, we found that the localization area always departs from its prior at the same time as the frequency, but its precision is generally better than that of the frequency, reaching $\Delta\Omega_{68}/\Delta\Omega_{100} \lesssim 0.1-1\%$ for all four sources. This behavior arises from the expected scaling relationships for these parameters: the precision on $\cos\theta$ and $\phi$ individually scales as (S/N)$^{-1}$, but the influence of the two parameters together causes the localization area to scale as (S/N)$^{-2}$ \cite{SV2010}. Finally, we note that 0.1$-$1\% of the entire sky (the constraint at (S/N)$_\Lambda \sim 20$) is an area of $\sim$40$-$400 deg$^2$, a region of significant size for EM follow-up efforts.

\textit{Chirp Mass} --- The chirp mass $\log_{10}\mathcal{M}$ breaks away from its prior slightly later than the sky location parameters, or not at all, depending on the source's GW frequency. For $f_{\rm GW}=$ 5 nHz (left column of \autoref{fig:qcw_par_order}), the chirp mass is effectively unconstrained, adhering to the full range of the prior across all time slices. For $f_{\rm GW}=$ 20 nHz (right column), the precision on the chirp mass decreases monotonically after it falls away from the prior, landing at $1\% \lesssim \Delta X_{68}/\Delta X_{100} \lesssim 10\%$ by the final time slice. At the end, the precision on $\log_{10}\mathcal{M}$ is similar to that of the $\log_{10}h_0$, $\cos\theta$, and $\phi$, or slightly outperforms these parameters in some cases. In terms of scaling, the chirp mass follows $\Delta X_{68}/\Delta X_{100} \propto$ (S/N)$^{-1}$.

The influence of GW frequency on the chirp mass precision is expected, as it is well-known that the ability to measure the binary chirp mass is contingent on the measurement of frequency evolution \cite{2002ApJ...575.1030T,CC2010,SV2010}. While typical PTA datasets are not long enough to see an appreciable ``chirp" in GW frequency, the necessary evolution can instead be found between the Earth and pulsar terms. Pulsar terms act as snapshots of the binary's orbital dynamics from $\sim$thousands of years before the Earth term snapshot. Consequently, the pulsar terms can provide constraints on the chirp mass if their frequencies are sufficiently different from the Earth term frequency. This difference can be achieved for high-frequency binaries, as they evolve more quickly, as in \autoref{eq:domega_dt}. The slow-evolving orbits of A5 and B5, therefore, do not induce enough separation between Earth and pulsar terms to assist in constraining the chirp mass. The opposite is true for the faster-evolving orbits of A20 and B20, whose pulsar terms have a larger difference from the Earth term, thereby providing information with which the chirp mass can be measured.

It should be noted that these pulsar term contributions derive entirely from frequency ``chirping" as described above. Binary parameter estimates can also be improved by measuring the phase interference between the Earth and pulsar terms, but this information is attained only when the pulsar distances are well-measured \cite{Lee2011,BP2012}. In our analyses, we do not benefit from the full power of the pulsar term, because i) the majority of pulsar distances are poorly measured (on the order of $\sim20\%$), and ii) we treat the pulsar phases as free parameters.

\textit{Inclination} --- The inclination angle $\cos\iota$ is usually the last parameter to receive constraints from the data, and its precision remains around $\Delta X_{68}/\Delta X_{100} > 10\%$ in all time slices. Similar to $\log_{10}h_0$, the precision on $\cos\iota$ appears slightly shallower than the expected (S/N)$^{-1}$ scaling, which is likely due to the injected face-on inclination.

When the binary's inclination is face-on, the measurement precision is expected to be poor for both the inclination itself and the strain. Towards edge-on inclinations, however, these parameters can be measured with better precision \cite{SV2010}. This is due to the fact that face-on binaries ($\iota=0$) create equal power in $h_+$ and $h_\times$ (see \autoref{eq:h_+} and \autoref{eq:h_x}), whereas inclinations departing from face-on create different amounts of power in the two GW polarizations, breaking the degeneracies between $\cos\iota$ and $\log_{10}h_0$, as well as parameters $\psi$ and $\Phi_0$.

In this work, we injected all CW signals face-on in order to maximize the strain and, consequently, the (S/N)$_{\rm opt}$. In terms of parameter precision, though, this choice represents a pessimistic scenario. An intermediate or edge-on inclination, maintaining the same signal strength throughout time slices, would likely rectify the plateau in precision seen for $\cos\iota$ and $\log_{10}h_0$ in \autoref{fig:qcw_par_order}. The parameters $\cos\theta$, $\phi$, $\log_{10}f_{\rm{GW}}$, and $\log_{10}\mathcal{M}$, on the other hand, are not strongly correlated with the inclination angle, so their precision should not change significantly with the choice of $\cos\iota$ \cite{SV2010}.

\textit{Other Parameters} --- Finally, we note that two CW parameters are not included in \autoref{fig:qcw_par_order}: the GW polarization angle $\psi$ and the initial binary phase $\Phi_0$. The posterior distributions of these parameters do change qualitatively across time slices, originally returning the prior and later showing increasingly more prominent peaks. At these later slices, the parameters' posteriors are also bimodal, owing to their degeneracy \cite{SV2010,2024PhRvL.132f1401C}. Even so, the posteriors always span the full range of the prior. As a result, these parameters would appear on \autoref{fig:qcw_par_order} much like $\log_{10}\mathcal{M}$ in the 5 nHz panels.

As mentioned previously, the measurement precision on $\psi$ and $\Phi_0$ is expected to be poorest for binaries with face-on inclination. Towards edge-on inclinations, the degeneracy between $\psi$ and $\Phi_0$ breaks due to the different $+$ and $\times$ contributions described above. Thus, in the face-on case, $\psi$ and $\Phi_0$ may not be precisely measured until PTA datasets have detected a CW signal with (S/N)$_\Lambda \gtrsim$~20. At different inclinations, however, these parameters may achieve more significant constraints than those shown here.

\textit{Frequency Dependence} --- The measurement precision attained in A20 and B20 outperforms that of A5 and B5 in regards to four parameters: $\log_{10}\mathcal{M}$, $\log_{10}f_{\rm{GW}}$, $\cos\theta$, and $\phi$. Again, the chirp mass can be measured more precisely for higher-frequency, faster-evolving binaries due to pulsar term contributions. The pulsar terms also influence the precision achieved for the GW frequency: at all time slices, A20 and B20 have better constrained frequencies owing to the greater frequency evolution between Earth and pulsar terms. For A5 and B5, the two terms have similar frequencies, and the CW signal is approximately a sum of two sinusoids of different phases. Lower-frequency binaries therefore have somewhat weaker frequency constraints.

The sky location parameters $\cos\theta$ and $\phi$ benefit slightly from higher-frequency signals, especially towards the final time slices. The fact that the pulsar terms supply information about the binary's sky location is not new; for instance, \citet{CC2010} found that pulsar terms not only disentangle the binary distance and chirp mass from the strain, as in \autoref{eq:h}, but also improve sky location measurements (see also \cite{Lee2011,BP2012,2023PhRvD.108l3535K}). P24 \cite{Petrov24} similarly showed that higher chirp masses induce larger separations between the Earth and pulsar terms and consequently achieve better sky constraints. More recently, \citet{2025arXiv250602819K} tested localization performance when one or two pulsars have precisely-measured distances, finding significant improvements when the source is located near or in between such pulsars. Pulsar terms also remove biases in sky location recovery, either when they are included in a full-signal analysis \cite{Zhu2016}, or when they are excluded from the analysis, i.e., Earth-term-only, but have frequencies significantly different from the Earth term frequency \cite{2025arXiv251204589G}.

This relationship between pulsar terms and the binary's sky location is exhibited clearly in \autoref{eq:t_p}; the time at which the GW passes the pulsar, $t_p$, encodes both the direction of GW propagation $\hat{\Omega}$ and the line of sight to the pulsar $\hat{p}$, meaning that the pulsar term is a function of the binary's sky location. Again, in our case, any localization assistance gained from the pulsar terms comes from some measurement of frequency evolution, rather than from ``phasing up the array" as explored in, e.g., \cite{Lee2011,BP2012}.

\textit{Sky Location Dependence} --- Although the order in which the CW parameters are constrained is fairly similar across all four simulation sets, locations A (top row of \autoref{fig:qcw_par_order}) and B (bottom row) differ in regards to the time slice (or (S/N)$_\Lambda$) at which the data become informative. In location B, the parameters pull away from the prior at earlier time slices; this separation is noticeable at the 10-year time slice, or (S/N)$_\Lambda$~$\sim$~7.1 and (S/N)$_\Lambda$~$\sim$~5.6 for B5 and B20, respectively. In location A, however, the parameters become constrained a few years later. For A20 these constraints kick in at 13-14 years, or (S/N)$_\Lambda$~$\gtrsim$~9.8, and for A5 at 14-17 years, or (S/N)$_\Lambda$~$\gtrsim$~10.5. We investigate this dependence on sky location further in the following section.

\subsection{Sky-dependent behavior}

\begin{figure}[!ht]
     \centering
     \includegraphics[width=0.48\textwidth]{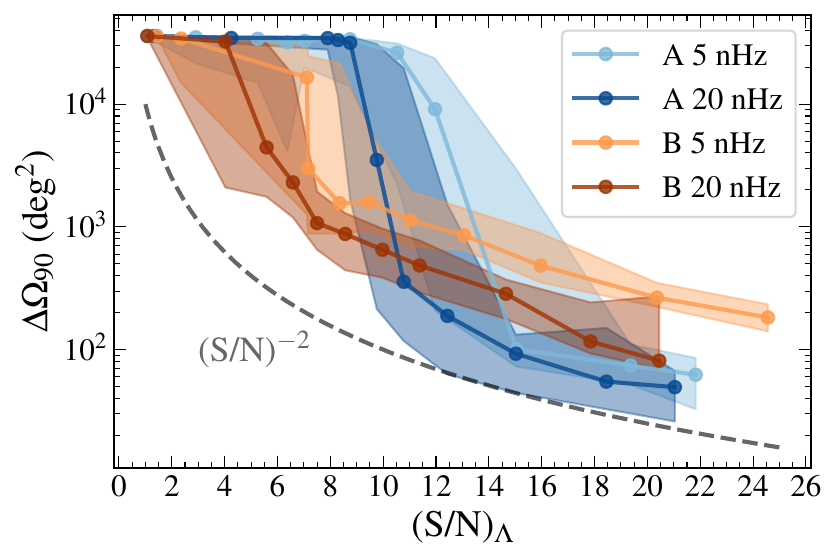}
     \caption{90\% credible area as a function of (S/N)$_{\Lambda}$. Solid lines show the median value of both (S/N)$_{\Lambda}$ and $\Delta\Omega_{90}$, while the shaded regions show the full range of values across all 10 realizations. Light and dark blue (orange) curves correspond to the A5 and A20 (B5 and B20) injections, respectively. The dashed gray line indicates the (S/N)$^{-2}$ scaling relation.}
     \label{fig:qcw_loc90}
\end{figure}

In \autoref{fig:qcw_loc90} we compare the constraints achieved between the four simulation sets more closely, specifically examining the localization area precision. For each simulation set we present the 90\% credible area $\Delta\Omega_{90}$, with solid lines indicating the median across realizations and shaded regions spanning the full range of values. Note that while \autoref{fig:qcw_par_order} shows the 68\% credible areas, \autoref{fig:qcw_loc90} instead shows the 90\% credible areas, as this standard is used across the GW spectrum. The time-evolving localization does not change drastically with the choice of credible level. Simulation sets A5 and A20 (B5 and B20) are shown in light and dark blue (orange), respectively. For reference, we also highlight the expected (S/N)$^{-2}$ scaling relation with a gray dashed line.

In contrast to \autoref{fig:qcw_par_order}, here we show the localization areas as a function of (S/N)$_\Lambda$ rather than time slice. Time slices are useful for tracking the evolving parameter precision within this study, but they are not widely applicable to existing PTA datasets and their current timespans. For example, the NANOGrav 15-year dataset cannot be directly compared to our 15-year time slice. Our 15-year slice contains a CW signal with (S/N)$_\Lambda$~$\gtrsim 11.5$, while the NANOGrav 15-year dataset did not find evidence for a CW signal \cite{NG15_cw}. Thus, subsequent time slices in our dataset (17-22 years) will not necessarily be predictive of what future NANOGrav datasets may observe. The more relevant quantity is (S/N)$_\Lambda$; once PTAs detect a CW signal with some level of (S/N)$_\Lambda$, this detection can be compared against our results to predict how parameter precision may evolve as the signal strength grows. From here onward we therefore compare the four simulation sets in terms of (S/N)$_\Lambda$ values.

The order in which the sources become localized, from lowest to highest (S/N)$_\Lambda$, is: B20, B5, A20, A5. Again as in \autoref{fig:qcw_par_order}, we can clearly see that sources in sky location B pull away from the prior first, or at lower (S/N)$_\Lambda$, compared to the sources in sky location A. Both B5 and B20 start to become constrained at (S/N)$_\Lambda$ $\sim 5-6$. The A20 localization area begins to improve only around (S/N)$_\Lambda$ $\sim 10$, yet by (S/N)$_\Lambda$ $\sim 11$ it has already surpassed both B sources. The same trend is generally seen for the A5 localization area, albeit at higher (S/N)$_\Lambda$; the area first declines around (S/N)$_\Lambda$ $\sim 11.6$ but by (S/N)$_\Lambda$ $\sim 15$ has become smaller than that of both B5 and B20. 

After each source is initially localized, it follows a gradual shrinking in localization area consistent with the expected (S/N)$^{-2}$ scaling \cite{SV2010}. In \autoref{fig:qcw_loc90} all four sources have reached this scaling-defined regime around (S/N)$_\Lambda$ $\gtrsim 15$. Here, the ordering of sources in terms of best to worst localization (or smallest to largest area) has changed. The new order is: A20, A5, B20, B5. 

The hierarchy of source localization at higher S/N is well in line with our expectations. Since location A is densely populated with nearby pulsars (see \autoref{fig:skymap}), we expect sources in this location to be better localized than sources in location B. This influence of pulsar proximity on CW localization has been identified previously, e.g., \cite{SV2010,BP2012,2018MNRAS.477.5447G,G19,Petrov24,2025arXiv250602819K}. We also expect higher-frequency sources to be somewhat better-localized than lower-frequency sources due to the influence of the pulsar terms, as discussed in Section \ref{sec:Res1}. Accordingly, in the high-S/N regime of \autoref{fig:qcw_loc90}, we see both A sources outperform the B sources, and A20 outperforms A5 while B20 outperforms B5.

The hierarchy of source localization at lower S/N, on the other hand, is not so straightforward. As expected, the 20 nHz sources still outperform their 5 nHz counterparts. However, contrary to the results at high S/N, location B is localized while location A is virtually unconstrained. We note that there is no time slice at which location B is denser with pulsars than location A; therefore, location B does not benefit from nearby pulsars at earlier time slices. These results point to the possibility that some factor other than pulsar proximity is influencing the initial localization of sources B5 and B20. We discuss and speculate on this idea in the next section.

\subsection{Pulsar term effects}

Thus far, we have noted the impact of S/N, pulsar proximity, and pulsar terms on sky localization. Because the S/N and pulsar proximity alone cannot explain the localization behavior of sources B5 and B20, we now turn to the contribution of pulsar terms. 

\begin{figure}[!ht]
     \centering
     \includegraphics[width=0.48\textwidth]{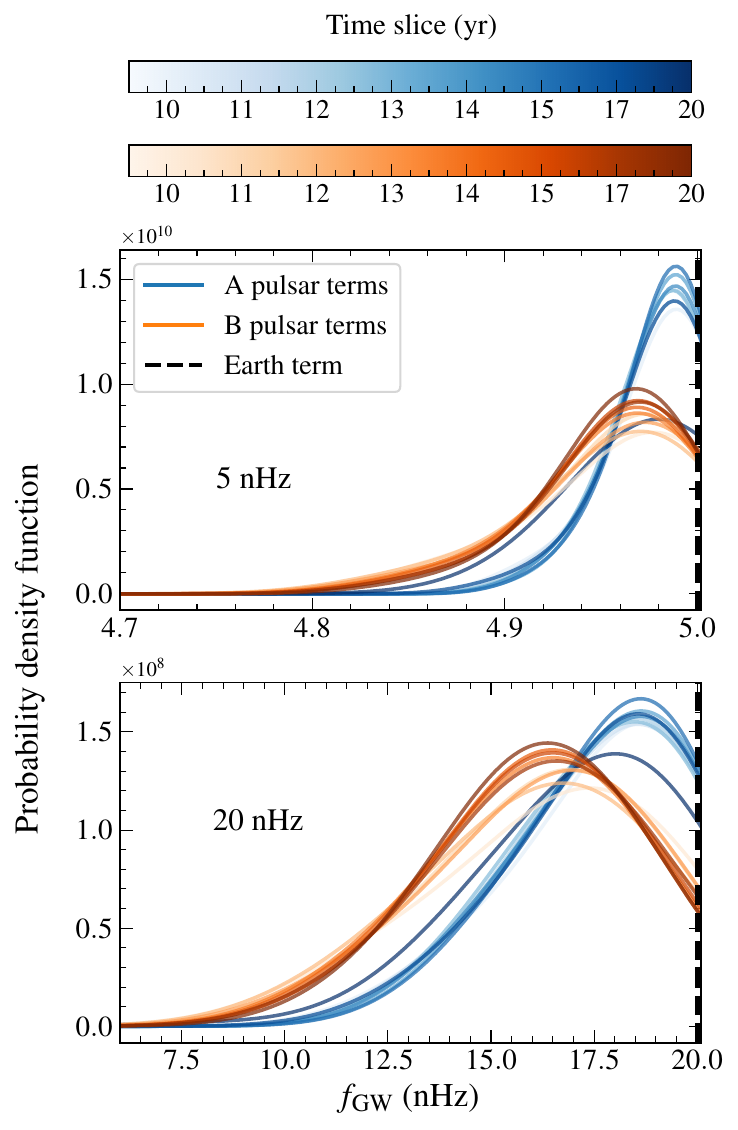}
     \caption{Distributions of pulsar term frequencies for all four injections. Sources with an Earth term frequency of 5 nHz are shown in the top panel, and those with an Earth term frequency of 20 nHz are shown in the bottom panel. The vertical dashed black line at the rightmost edge of each panel indicates the injected Earth term frequency. Pulsar term frequency distributions for locations A and B are shown with blue and orange curves, respectively. Lighter-colored curves show distributions for earlier time slices, while progressively darker-colored curves show distributions for later time slices. Note that the first (5-year) and last (22-year) time slices are excluded here; in the first slice the array contains only 5 pulsars, and the last slice contains the same number of pulsars as the 20-year slice.}
     \label{fig:pterm_freqs}
\end{figure}

\begin{figure*}[!ht]
     \centering
     \includegraphics[width=0.98\textwidth]{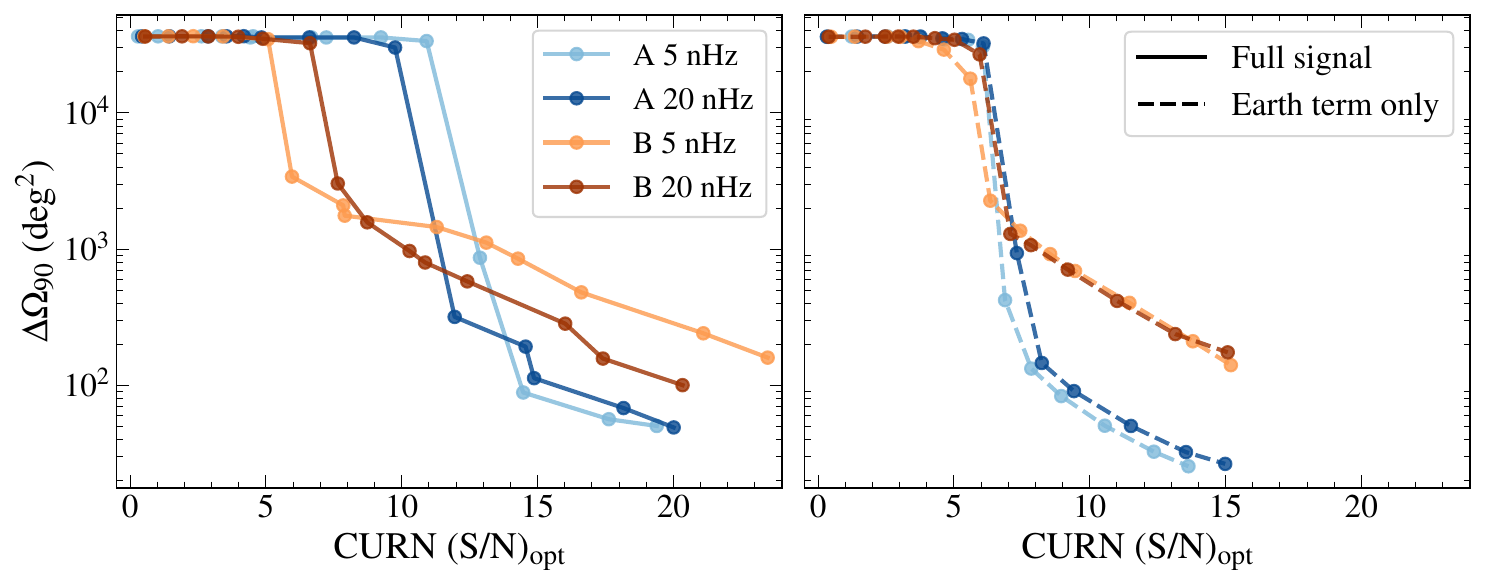}
     \caption{90\% credible areas for noiseless CW injections, keeping noise parameters fixed in the covariance matrix of the analyses. Areas are shown as a function of (S/N)$_{\rm{opt}}$ calculated with a CURN model. Color scheme is the same as in \autoref{fig:qcw_loc90}, with full-signal injection+model shown in the left panel (solid lines) and Earth-term-only injection+model shown in the right panel (dashed lines). The binary parameters in the Earth-term-only injections are identical to those used in the full-signal injections.}
     \label{fig:fs_vs_eto}
\end{figure*}

To investigate the role that pulsar terms may play in the localization of our sources, we calculate the pulsar term frequency for each pulsar in the array via \autoref{eq:w_p}. For any given pulsar, the pulsar term frequency will depend on its distance, its angular separation from the source on the sky, and the source's frequency. The chirp mass in our simulations is always fixed to $\mathcal{M}=10^9 M_{\odot}$ and consequently does not play a role in changing the pulsar term frequencies from injection to injection.

In \autoref{fig:pterm_freqs} we show the distribution of pulsar term frequencies for all four sources and for different time slices. The 5 nHz (20 nHz) sources are shown in the top (bottom) panel, and each panel shows the injected Earth term frequency with a vertical black dashed line at the rightmost edge. Distributions in blue and orange correspond to sources in locations A and B, respectively. We plot the distributions starting with the 7-year time slice and ending with the 20-year time slice; progressively darker-colored curves indicate later time slices. With each subsequent slice, the distributions change only due to the addition of new pulsars; each pulsar term frequency does not change from one time slice to another.

Across all time slices in both panels, the pulsar terms for location B are more widely separated from the Earth term compared to that of location A. Considering the distribution of pulsars on the sky in \autoref{fig:skymap}, this is expected. Location B has fairly large angular separations from the array's pulsars. Because the pulsar term depends on $\hat{\Omega} \cdot \hat{p} = -\cos\mu$ where $\mu$ is the angular separation, location B's large angular separations result in large differences between the Earth and pulsar term frequencies. Conversely, the high density of pulsars around location A results in small angular separations between the source and the pulsars. The pulsar terms in location A therefore do not achieve as great a separation from the Earth term as in location B. 

The separation in frequencies can generally be compared in terms of frequency bins. Note that the timespan $T$ of each time-sliced dataset is different, hence the frequency bins with bin-centers $f_i = i/T$ differ as well. The bins become more and more narrow with each successive slice, starting with a bin width of $1/(5\rm{yr}) = 6.34$ nHz and ending with a bin width of $1/(22\rm{yr}) = 1.44$ nHz. For A5 and B5 in the top panel of \autoref{fig:pterm_freqs}, across all slices, the pulsar terms lie in the same bin as the Earth term. More specifically, this frequency bin spans 4.75-7.92 nHz for the 7-year slice and 4.66-6.52 nHz for the 20-year slice. For A20 and B20 in the bottom panel, there is always some number of pulsar terms lying outside the Earth term's bin. At the 7-year slice, A20 (B20) has 3 (6) of 20 pulsars lying outside the 17.4-20.6 nHz bin. At the 20-year slice, A20 (B20) has 109 (115) of 116 pulsars lying outside the 19.6-21.4 nHz bin.

We note, however, that Bayesian CW analyses are not limited to Fourier-bin resolution in the way that GWB analyses are. Consequently, any pronounced separation between Earth and pulsar terms, even if they remain within the same frequency bin, can aid in source characterization. A greater offset in pulsar term frequencies will improve sky localization, as there is more distinct information with which the binary's sky location can be solved \cite{CC2010,Lee2011,2025arXiv251204589G}. Thus, sources in location B are at a slight advantage in that they receive more pulsar term contribution to parameter estimation.

\subsubsection{Full-signal versus Earth-term-only}

To isolate pulsar term effects on sources A and B, we compare a full-signal injection+recovery (including both Earth and pulsar terms) to that of an Earth-term-only injection+recovery. The results are shown in \autoref{fig:fs_vs_eto}. The four sources have the same colors as in \autoref{fig:qcw_loc90}, while solid and dashed lines correspond to the full signal (left panel) and Earth-term-only (right panel), respectively. The injections and analyses are different from those shown in \autoref{fig:qcw_loc90}; here we inject only a CW signal, excluding the injection of white noise, intrinsic pulsar red noise, and the GWB. We instead account for these noise processes by fixing their parameters in the PTA covariance matrix during the MCMC analysis. Performing the analysis in this way allows for an effectively noise-realization-averaged result \cite{Nissanke2010,Cornish2010}.

Note that since we are working with noiseless injections, localization areas in \autoref{fig:fs_vs_eto} are shown as a function of (S/N)$_{\rm{opt}}$ rather than (S/N)$_\Lambda$. Unlike the (S/N)$_\Lambda$ values, which assumed an HD-correlated GWB, the (S/N)$_{\rm{opt}}$ values assume a CURN model for the GWB. This choice is sensible given that \texttt{QuickCW} uses a CURN model, and in this test we are simply interested in comparing the localization performance of the four sources. We note also that we keep the source parameters identical between the full-signal and Earth-term-only injections; the Earth-term-only sources therefore do not reach as high a (S/N)$_{\rm{opt}}$ as the full-signal sources. The difference in (S/N)$_{\rm{opt}}$ is not unexpected, as the pulsar terms contribute to the signal's strength \cite{CC2010,Lee2011,Zhu2016}.

First, we compare the full-signal results in \autoref{fig:fs_vs_eto} to those in \autoref{fig:qcw_loc90}. Overall, the localization behavior is very similar, with one minor difference being that B5 is now localized before B20. This change may arise from fixing all CURN and RN parameters in the noise covariance matrix. When sampling over all parameters together, low-frequency CW signals may be harder to disentangle from red noise processes which also manifest at low frequencies. This is especially true for early time slices, in which B5 has a small number of GW cycles. Fixing the noise processes could therefore allow for easier convergence on B5's signal. Similar conflation issues have also been seen in, e.g., searches for periodicity in quasar light curves \cite{2016MNRAS.461.3145V}. Ultimately, though, the (S/N)$_{\rm{opt}}$ range at which B5 and B20 become constrained is comparable.

Another small change from \autoref{fig:qcw_loc90} is that once A5 is localized, it very narrowly outperforms A20. Of course, fixing the noise parameters does not benefit A5 like it does B5, i.e., A5 does not become localized before A20. It is possible that A5's recovery is rather more limited by interference between Earth and pulsar terms than by competition with red noise. From \autoref{fig:pterm_freqs}, for instance, we see that A5's pulsar term frequencies have the smallest separation from the Earth term frequency compared to the other three sources. However, once A5's signal is detected, the fixing of RN and CURN parameters---i.e., removing potential conflation---may aid it in achieving slightly better localization than A20 at high (S/N)$_{\rm{opt}}$.

Overall, the general hierarchy of B sources versus A sources remains the same in \autoref{fig:fs_vs_eto}. B5 and B20 are again localized at lower (S/N)$_{\rm{opt}}$ than A5 and A20, while A5 and A20 outperform B5 and B20 at higher (S/N)$_{\rm{opt}}$. This hierarchy, then, is not likely being influenced by noise.

Now we compare the Earth-term-only results in \autoref{fig:fs_vs_eto} to those of the full signal. Having removed the influence of pulsar terms, the localization in the Earth-term-only case is markedly different; all four sources become localized around (S/N)$_{\rm{opt}}~\sim~6-7$, after which they follow the expected (S/N)$^{-2}$ scaling. As in the full-signal case, A5 and A20 experience a steeper drop in localization area compared to B5 and B20. Following this drop, sources in location A are always better localized than those in location B, in line with the expectation that sources surrounded by many nearby pulsars achieve better localization \cite{SV2010,BP2012,2018MNRAS.477.5447G,G19,Petrov24,2025arXiv250602819K}.

Clearly, the localization of a GW source can be more or less impacted by pulsar term contributions, depending on the source's sky location relative to the array. When pulsar terms are included, sources A and B are localized at different S/N regimes, with B sources particularly benefiting from pulsar terms. This result is fairly intuitive for source B20, which, among the four sources, has the greatest difference between pulsar term and Earth term frequencies (see \autoref{fig:pterm_freqs}). Source B5 is somewhat less straightforward. \autoref{fig:pterm_freqs} shows that B5's frequency evolution between Earth and pulsar terms (of order $\lesssim 0.1$ nHz) is larger than that of A5, yet smaller than that of A20 (of order $\sim$ one to a few nHz). The ``earlier" localization of B5 is therefore reasonable compared to A5, but necessitates further inspection in relation to A20.

\subsubsection{Pulsar distance and phase parameters}

Pulsar terms can be better constrained, and in turn the source localization can be improved, if precise pulsar distance measurements are obtained \cite{BP2012,2023PhRvD.108l3535K,2025arXiv250602819K}. For a pulsar's distance to be classified as accurately measured, the precision should generally be smaller than the gravitational wavelength $\lambda_{\rm{GW}} = c/f_{\rm{GW}}$ (typically of order $\sim$1 pc). More specifically, the distance should be known well relative to the quantity $\lambda_{\rm{GW}} / (1+\hat{\Omega}\cdot\hat{p})$, which includes the angular separation between the source and the pulsar \cite{BP2012}. Such high quality measurements are difficult to achieve for every pulsar in the array, but even just one or two well-known pulsar distances can improve localization performance across the celestial sphere \cite{2025arXiv250602819K}. Measurements approaching, but not yet satisfying, the $\lambda_{\rm{GW}} / (1+\hat{\Omega}\cdot\hat{p})$ criterion can also be of increased benefit.

\begin{figure}[!ht]
     \centering
     \includegraphics[width=0.48\textwidth]{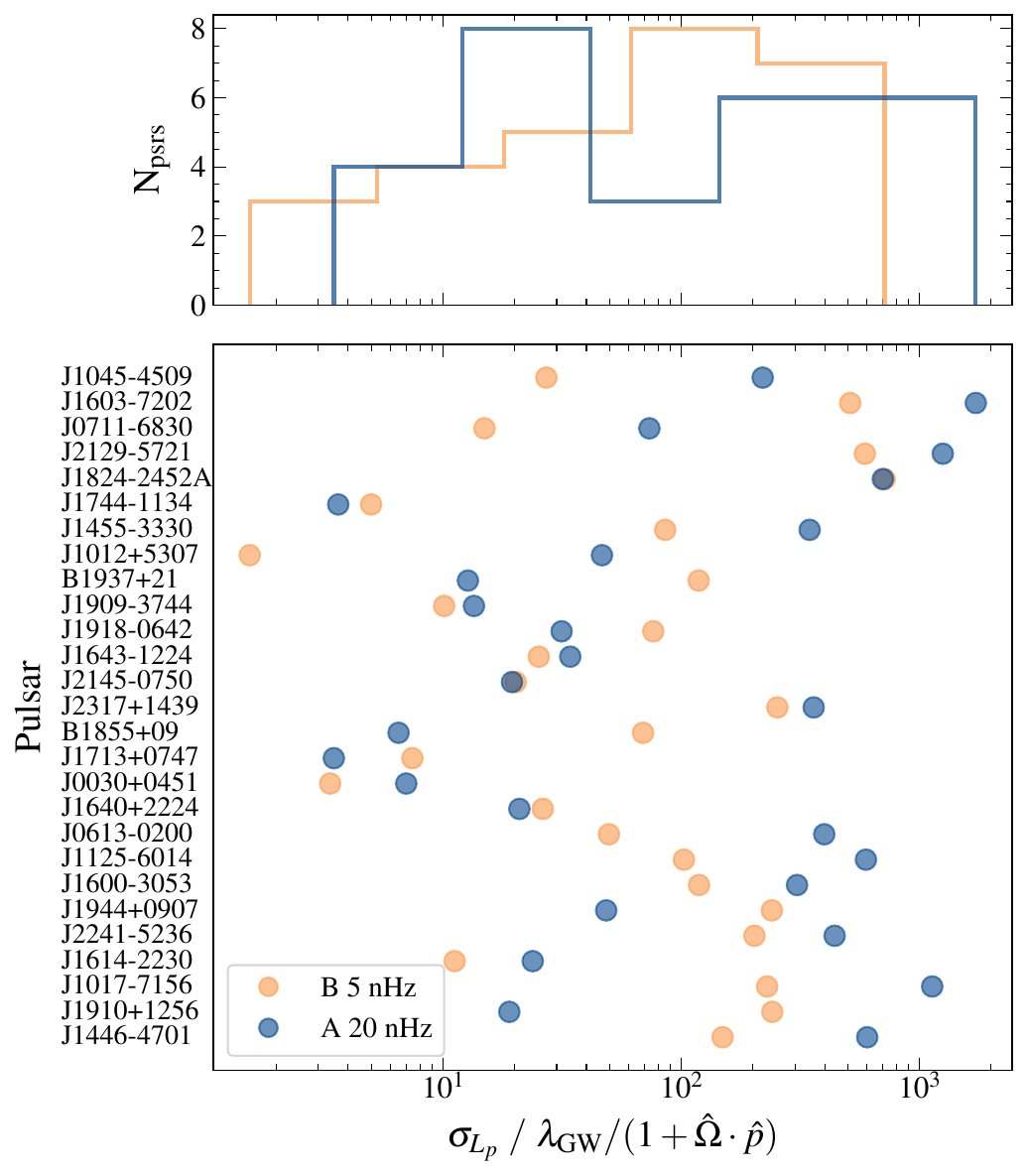}
     \caption{Distribution of pulsar distance uncertainties $\sigma_{L_p}$ for the 27 pulsars in the 10-year slice, according to how well the distances are known relative to $\lambda_{\rm{GW}}/(1+\hat{\Omega}\cdot\hat{p})$. From top to bottom, the pulsars are ordered by the slice in which they are added to the array, i.e., the 5-year (7-year) slice is comprised of the top 5 (20) pulsars. Sources A20 and B5 are shown in dark blue and light orange, respectively. Smaller values indicate more precisely-measured distances, with a value $\leq 1$ designating the threshold for an accurately-known distance. (In this figure, no pulsars meet this criterion.)}
     \label{fig:bpcrit}
\end{figure}

\begin{figure*}[!ht]
     \centering
     \includegraphics[width=0.98\textwidth]{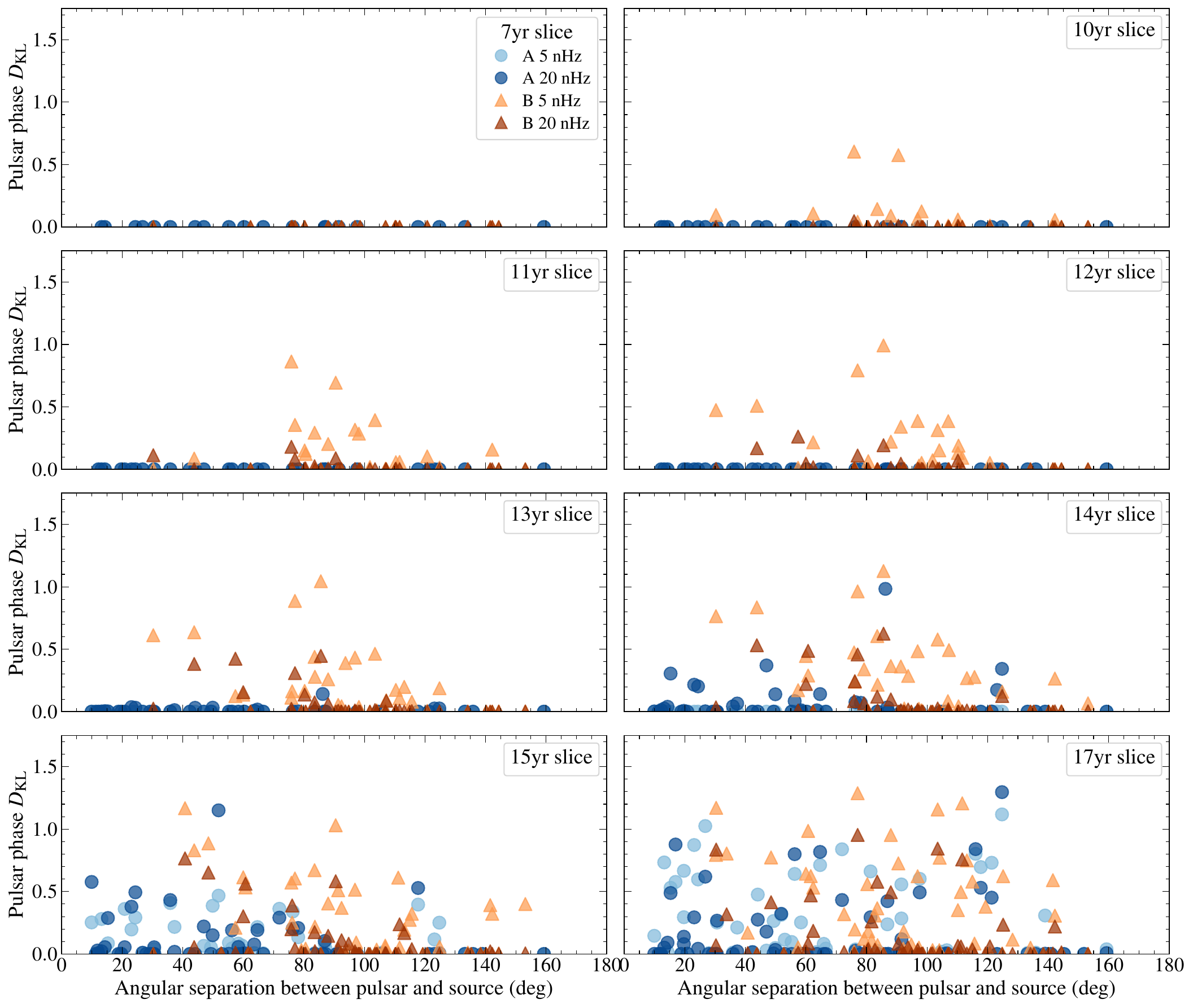}
     \caption{Kullback–Leibler (KL) divergence computed between the pulsar phase posterior and prior for each pulsar in the array, plotted as a function of the pulsar's angular separation from the source. The four noiseless injections are shown, again with posteriors obtained by fixing the noise parameters in the PTA covariance matrix; hence, these results correspond to the same datasets and analyses shown in the left panel of \autoref{fig:fs_vs_eto}. Sources in locations A and B are shown with circle and triangle markers, respectively, and all are colored in the same manner as in previous figures. For each source, the initial increase in pulsar phase $D_{\rm{KL}}$ corresponds to the initial improvement in localization area in \autoref{fig:fs_vs_eto}.}
     \label{fig:kldiv}
\end{figure*}

For GW sources with lower $f_{\rm{GW}}$, or longer $\lambda_{\rm{GW}}$, the pulsar distance precision requirement is less stringent. Comparing sources B5 and A20, B5 has a longer $\lambda_{\rm{GW}}$ ($\sim1.9$ pc) than A20 ($\sim0.5$ pc). B5 is therefore at an advantage in this context, since more pulsars may have distance precisions $\lesssim 1.9$ pc. To this point, we show a more thorough comparison of pulsar distance measurements in \autoref{fig:bpcrit}, where we plot the ratio of the pulsar distance uncertainty $\sigma_{L_p}$ over the quantity $\lambda_{\rm{GW}} / (1+\hat{\Omega}\cdot\hat{p})$. A smaller ratio signifies a more precisely-measured distance, and a ratio $\leq 1$ satisfies the criterion for an accurately-known distance. We include only the 27 pulsars from the 10-year time slice, at which B5's parameters first become constrained. The top panel shows the distribution of ratios for all pulsars together, while the bottom panel shows each pulsar's ratio individually.

Although no pulsars meet the criterion for either source B5 (light orange) or A5 (dark blue), the top panel shows that B5's distribution of ratios is shifted towards smaller values than that of A20. In the bottom panel, we find that more than half of the pulsars (17 of 27) have a smaller ratio $\sigma_{L_p}$~$/$~$\lambda_{\rm{GW}}/(1+\hat{\Omega}\cdot\hat{p})$ for B5 than for A20. With more pulsars approaching the accurately-known distance criterion, the pulsar terms may be contributing more significantly to B5's parameter estimation, resulting in constraints that emerge earlier, or at lower S/N.

As a further test of pulsar term influence, we examine the pulsar phase parameters. We compute the information gain in these parameters using the Kullback-Leibler (KL) divergence $D_{\rm{KL}}$ \cite{KLdivergence}, a statistic that quantifies the difference between two probability distributions. In this case, we are interested in the difference between the pulsar phase posterior and prior, which we calculate for each pulsar individually. More concretely, we compare each marginalized pulsar phase posterior from the MCMC chain against a uniform prior between 0 and $2\pi$. The two distributions are identical if $D_{\rm{KL}}=0$, while larger values mean that the distributions are more distinct from one another. Put simply, larger $D_{\rm{KL}}$ implies greater information gain from the data. \autoref{fig:kldiv} shows $D_{\rm{KL}}$ values for every pulsar phase parameter as a function of that pulsar's angular separation from the source. Sources in locations A and B are shown with circle and triangle markers, respectively, with the same coloring as in previous figures.

We note that the calculations in \autoref{fig:kldiv} were done on the noiseless injection posteriors and consequently follow the trends seen in the left panel of \autoref{fig:fs_vs_eto}. For each of the four sources, we find an increase in pulsar phase $D_{\rm{KL}}$ at the same time that the full-signal localization area in \autoref{fig:fs_vs_eto} improves. This occurs at the 10-year slice for B5, 11-12-year slices for B20,  13-14-year slices for A20, and 15-year slice for A5. In \autoref{fig:kldiv} as in \autoref{fig:fs_vs_eto}, B5 shows parameter constraints slightly before B20; again, this may be due to the fact that, since the noise parameters are fixed in the PTA covariance matrix, conflation between red noise and the CW signal during sampling has been mitigated. The KL divergence results also do not change drastically for the noisy injections -- the only difference is that both B5 and B20 see a rise in pulsar phase $D_{\rm{KL}}$ at the 10-year slice, coinciding with the initial parameter improvements in \autoref{fig:qcw_par_order} and \autoref{fig:qcw_loc90}.

We can conclude, then, that the data become informative simultaneously for global CW parameters such as $(\theta,\phi)$, $f_{\rm{GW}}$, and $h_0$, as well as several pulsar phases $\Phi_p$. This provides additional support that B5, and both B sources in general, are constrained at lower S/N due to pulsar term influence. These pulsar term gains are fairly moderate; more drastic improvements could be achieved by leveraging pulsar term phasing information, as discussed previously. However, both in this work and in real PTA analyses, obtaining such improvements is currently unfeasible without precise distance measurements for the majority of pulsars.

\subsection{\label{sec:othertests}Additional tests}

In this section, we discuss two additional tests which produced inconclusive results. In these tests, we ran identical analyses to those shown in \autoref{fig:fs_vs_eto}, but with a static configuration of pulsars. The results are shown in \autoref{fig:staticPTA_tests} of the Appendix.

In the first test, we took only the subset of 20 pulsars from the 7-year time slice and reran the analyses for each successive slice. The injected CW source parameters were unchanged. This test was conducted in order to reduce the number of factors affecting the (S/N)$_{\rm{opt}}$ from one time slice to the next; here the increase in (S/N)$_{\rm{opt}}$ comes from the increase in timespan alone, using pulsars that accumulate fairly long baselines by the final dataset. Ultimately, we found that the hierarchy of sources remained the same at both low and high (S/N)$_{\rm{opt}}$ regimes. While this result is generally uninformative, it suggests that this specific set of 20 pulsars aids in the B sources' parameter estimation.

In the second test, our fixed PTA configuration included all 116 pulsars and the full 22-year timespan. This time, rather than varying (S/N)$_{\rm{opt}}$ by adding pulsars and observations, we instead modified the luminosity distance to the source. We note that by adjusting (S/N)$_{\rm{opt}}$ in this way, only the amplitude of the signal is affected, while the signal structure (including that of the pulsar terms) is no different from our original simulations. 

This test was performed for a couple reasons. First, compared to the static 20-pulsar test above, the 116-pulsar configuration is somewhat more isotropic over the celestial sphere. Second, our use of the full 22-year dataset ensures an adequate number of GW cycles. 
Earlier time slices cover fewer wave cycles; this is particularly problematic for the 5 nHz sources, which have a cycle duration of $1/$(5 nHz)~$\sim$~6 years. For sufficiently long datasets, we expect that (S/N)$_{\rm{opt}} \propto \sqrt{T}$ \cite{2014ApJ...794..141A}. For datasets with a small number of wave cycles, however, this relationship may not hold, since $f_{\rm GW}$ becomes comparable to the time over which we integrate the inner product in \autoref{eq:snr_opt}. Parameter measurements may therefore scale nontrivially with (S/N)$_{\rm{opt}}$, until the behavior saturates at large times. In brief, by simulating a 22-year dataset with varied luminosity distances, we achieve a gradation in signal strength which scales trivially, (S/N)$_{\rm{opt}} \propto 1/d_L$, while avoiding any non-monotonic scaling arising from integration over few GW cycles.

From this test, we found that sources A20, B5, and B20 were all constrained around (S/N)$_{\rm{opt}} \sim 10$, and A5 at (S/N)$_{\rm{opt}} \sim 11.5$. Once constrained, the sources follow the same hierarchy as before, with A sources outperforming B sources. The fact that three sources emerged in the same (S/N)$_{\rm{opt}}$ regime (and the fourth not far behind) could imply that pulsar term effects related to sky location matter less for larger arrays. Although the 116-pulsar array is not uniformly distributed on the sky, it nonetheless more closely resembles an isotropic distribution. As a result, the spread of angular separations between pulsars and GW source become comparable for sky locations A and B. Coupled with this, the influence of PTA geometry dominates in this case because most pulsars have poor distance precision.

Compare this to the 20-pulsar PTA in the previous test, which is much more anisotropic, producing very different distributions of pulsar separations for locations A and B. Furthermore, that array has a larger fraction of decently-measured pulsar distances. We illustrate this in \autoref{fig:pdist_uncs} of the Appendix, which shows the distribution of pulsar distance uncertainties for the 20-pulsar (light gray) and 116-pulsar (dark gray) arrays. While both sets have the largest density of pulsars at the $\sim20\%$ level, the surrounding bins are noticeably different. For the 20-pulsar array, smaller \% uncertainties (higher distance precisions) rival those in the $20\%$ bin. Thus, pulsar terms have more influence over the source characterization. For the 116-pulsar array, however, the density of pulsars in the $20\%$ bin far outweighs that of pulsars with well-measured distances. In this heap of poor-distance-estimate pulsars, the source characterization is guided by pulsar proximity on the sky \cite{BP2012}.

\section{\label{sec:Discuss}Discussion}

\subsection{\label{sec:D1}Parameter estimation in other studies}

In this work, we use a PTA configuration identical to that of P24 \cite{Petrov24}, and inject CW signals into similar sky locations. P24 injected CWs into real galaxy locations, three of which are in a region like location A (near many pulsars) and another three of which are very close to location B (devoid of nearby pulsars). Two datasets were created for each galaxy location, one with an injected signal of (S/N)$_{\rm{opt}}=8$ and the other with (S/N)$_{\rm{opt}}=15$, and all had $f_{\rm GW} =$~20 nHz. In the following we compare P24's localization results to that of our A20 and B20 sources. 

At (S/N)$_{\rm{opt}}=15$ in \autoref{fig:fs_vs_eto}, our findings match that of Fig.~3 in P24. In both studies, we find that sources near location A achieve smaller localization areas than those near location B, thanks to the abundance of nearby pulsars. We generally see the same behavior in the noisy simulations in \autoref{fig:qcw_loc90}. At (S/N)$_{\rm{opt}}=8$, our findings differ somewhat from P24. In this paper, we find that B20 is constrained by (S/N)$_{\rm{opt}}=8$, but A20 is not. In Fig.~4 of P24, the opposite is true: all A-like sources are constrained, while all B-like sources are not, with 90\% credible areas spanning a majority of the sky.

These discrepancies likely stem from the different calibrations of CW (S/N)$_{\rm{opt}}$. Here, the (S/N)$_{\rm{opt}}$ grows through the addition of timing observations and pulsars. In P24, we obtained the desired (S/N)$_{\rm{opt}}$ by tuning the binary's chirp mass, meaning that the signal structure necessarily changes from (S/N)$_{\rm{opt}}=8$ to (S/N)$_{\rm{opt}}=15$ for the same source. Adjusting the signal in this way can create fluctuations in parameter estimation performance, as the chirp mass affects the pulsar term evolution, and in turn the summation of Earth and pulsar terms may constructively or destructively interfere. These fluctuations persist even for signals of fixed (S/N)$_{\rm{opt}}$, when binary evolution is minimal (see Fig.~7 of P24). Beyond this, the difference in injected locations may play a role. Specifically, our location A may not be directly comparable to P24's A-type locations. Despite being close in R.A. and near many pulsars, location A in this study sits at $\delta=+22^{\circ}$, while in P24 the injected locations range from $\delta\sim-40^{\circ}$ to $\delta\sim-27^{\circ}$. Finally, here we inject a louder GWB, with $A_{\rm GWB}=2.4\times10^{-15}$ based on NANOGrav's 15-year dataset \cite{NG15_gwb}, compared to P24's $A_{\rm GWB}=1.92\times10^{-15}$ based on NANOGrav's 12.5-year dataset \cite{NG12p5_gwb}. 

Our results can also be compared with Fig.~3 of \citet{G19} (hereafter G19), which shows localization areas for sources in three different sky locations. G19's hierarchy of sources at (S/N)$_{\rm{opt}} = 7$ resembles ours; that is, the localization of their A-like source is generally worse than that of their B-like source. By (S/N)$_{\rm{opt}} = 10$, the two sources have switched in localization performance, akin to the switch in source hierarchy shown in this paper.

It is interesting that in G19 this behavior persists despite their use of a null-stream analysis, which omits the pulsar terms. In our work, we find some evidence that location B initially overpowers location A due to pulsar terms contributions. Conversely, G19 do not model pulsar terms. They instead attribute the change in localization to the increase in triangulation power that comes from an increase in individual pulsar (S/N)$_{\rm{opt}}$. When calculating per-pulsar (S/N)$_{\rm{opt}}$ for our simulations, we find some marginal evidence in favor of this conclusion. For example, in earlier time slices, our B sources have a larger $N_{\rm pulsar}$ contributing to 90\% of the total (S/N)$_{\rm{opt}}$, hence more pulsars with which to localize; our A sources have fewer pulsars contributing to the total (S/N)$_{\rm{opt}}$, and may therefore have weaker localizing power. Overall, though, some caution should be given to direct comparison between G19 and this work, as both the PTA configuration and the injected source properties are different.

\subsection{\label{sec:D2}Realistic GW backgrounds and PTA datasets}

As mentioned in Section \ref{sec:Analyses}, our analyses model the GWB without HD spatial correlations. This results in a slight mismatch between our simulations and analyses, and requires a check of potential biases in the recovered parameters. For a visualization of parameter recovery accuracy, we refer the reader to Fig.~5 of Paper I.

We find some bias in the estimated GWB parameters, systematically recovering a shallower spectral index $\gamma_{\rm{GWB}}$ and higher amplitude $A_{\rm{GWB}}$ than injected. Previous studies have seen similar GWB biases when the model is mis-specified, e.g., applying a GWB-only model on a simulated GWB+CW dataset \cite{2023ApJ...959....9B,2024A&A...683A.201V,2024arXiv240721105F}. The GWB parameters can also be affected by spectral leakage: a finite observation window will produce frequency correlations, which, if not accounted for in the noise covariance matrix, can bias GWB recovery towards lower spectral indices and higher amplitudes. This effect becomes more important when more pulsars are analyzed \cite{2025arXiv250613866C}. In our analyses, too, we generally find that this bias is slightly exacerbated for later time slices (i.e., longer datasets with more pulsars).

The CW signal, on the other hand, generally shows unbiased recovery once the parameters are constrained. One exception is the CW strain $\log_{10}h_0$, which is often overestimated. This bias does not seem to relate to the bias in GWB parameters, however, as it persists in both our noisy and noiseless simulations. Instead, it has more to do with the recovery of $\log_{10}h_0$ together with the inclination angle $\cos\iota$. Our sources are injected with a face-on inclination, i.e., $\cos\iota=1$, and this being one of the prior boundaries, the sampler may struggle to sample this region of parameter space effectively. Alternatively, the sampler explores more edge-on inclinations ($\cos\iota<1$) while preserving the overall signal amplitude by increasing $\log_{10}h_0$ (see \autoref{eq:h_+} and \autoref{eq:h_x}). Accordingly, we find that when the $\log_{10}h_0$ posterior is biased higher, the $\cos\iota$ posterior is biased more edge-on. This degeneracy is well known and has been seen in the literature previously, e.g., \cite{Emiko25}.

In the future, additional work may be needed to investigate the evolution of CW parameter estimation when using an HD model as compared to a CURN model. Several studies note that obtaining confident, simultaneous recovery of all CW and GWB parameters may require modeling an HD-correlated GWB \cite{2023ApJ...959....9B,2024arXiv240721105F,2024A&A...683A.201V}. Aside from accurate recovery, searches for CWs in real PTA datasets have noted some covariance and competition between the HD and CW models \cite{NG15_cw,EPTA_InPTA_DR2_cw}. For example, searches for individual binaries in NANOGrav's 15-year data set produced Bayes factors that favored a CURN+CW model over CURN alone, yet favored HD alone over an HD+CW model \cite{NG15_cw}. Using the full HD+CW model on our simulations, it would be interesting to explore whether the parameter precision changes at earlier slices, where the CW signal is first emerging. This would likely impact the 5 nHz sources more than the 20 nHz sources, since the GWB is strongest at lower frequencies.

However, modeling the full complexity of an HD-correlated GWB together with a CW source is computationally demanding. While there are less costly techniques for incorporating HD correlations, e.g., sample reweighting \cite{Hourihane23}, this approach was likewise inefficient when applied to our analyses. In short, this technique can be ineffective when the CURN+CW and HD+CW posteriors do not overlap sufficiently \cite{Hourihane23,NG15_cw}. We refer the reader to Paper I for further details about reweighting and its inadequacy for our work. Beyond reweighting, new techniques for HD+CW modeling are vital, especially as PTA datasets grow larger in number of observations and pulsars; Fourier basis methods \cite{2024arXiv241213379G} and GPU accelerations are a couple such promising developments.

Finally, there are other aspects of our simulations and analyses that could benefit from a further dose of realism. Some features of our PTA are conservative, e.g., the pulsar timing precision is kept constant across time slices, while in real PTA datasets the precision typically improves the longer a pulsar is observed. At the same time, some aspects of our simulations are idealized. For example, the injection of a Gaussian GWB is simplistic and does not accurately reflect how overlapping GW signals will manifest in PTA data. A more realistic approach would be to simulate an entire population of SMBHBs, injecting CW signals from all systems into one dataset. Alternatively, one can inject the loudest 100 sources in each frequency bin, plus an isotropic GWB resulting from the remaining sources \cite{BCK22_realgwbs,Emiko24}. We also recover only one CW signal at a time; in reality, we may begin to resolve multiple CW signals simultaneously, necessitating a trans-dimensional signal model (e.g., \cite{2020CQGra..37m5011B}). For additional discussion on these points and other caveats, see Section V of Paper I.

\section{\label{sec:Conclus}Conclusions}

In this work, we studied CW parameter estimation across time-sliced datasets, tracking the evolution of measurement precision as the timespan, number of pulsars, and S/N increase concurrently. We find that the GW frequency and strain generally become constrained first, closely followed by the sky location parameters. All four parameters tend to pull away from their respective priors at a similar time or S/N. Consequently, when PTAs begin to detect a CW signal, we may see simultaneous constraints on the GW frequency, strain, and sky location, albeit at varying degrees of precision. 

The gain in precision for these parameters can occur after an interval of data collection as little as one year (during which new pulsars with $\geq$ 3 years of observations are also added). In reality, each new PTA dataset typically contains an additional $\gtrsim$ 2 years of data compared to the last, as well as some increase in the number of pulsars being monitored. Therefore, it is possible that one PTA dataset may not show evidence for a CW signal, but the next may suddenly show constraints similar to those exhibited here. This possibility mirrors the results in Paper I, where Bayes factors abruptly increased from one dataset to the next.

The chirp mass and inclination angle become constrained later, or at higher S/N. The precision on the chirp mass improves across time slices only if the source has a relatively high GW frequency; otherwise, the chirp mass cannot be constrained. The inclination angle, for face-on binaries, gradually improves in precision but is consistently the worst-measured parameter. The GW polarization angle and orbital phase are unconstrained relative to their priors, but their posteriors show qualitative changes across slices. For non-face-on binaries, the measurement precision on the GW strain, inclination, GW polarization, and initial phase would likely outperform the precision seen here.

We injected sources at two different sky locations, A (surrounded by many pulsars) and B (placed in a void of pulsars), as well as two different GW frequencies, 5 nHz (slowly-evolving) and 20 nHz (faster-evolving). Across these injections, we find that the measurement precision depends on the source's frequency and sky location. At fixed S/N, faster-evolving (higher frequency) binaries achieve better precision on the frequency, chirp mass, and sky location, due to information contributed by the pulsar terms. We also find that the B sources are better constrained at earlier time slices, or lower S/N, while the A sources are better constrained at later time slices. Despite the fact that sources A5 and A20 lag behind B5 and B20 in parameter estimation, the A sources experience a precipitous improvement in precision and ultimately outperform the B sources in the high-S/N regime.

When a CW source is detected by PTAs, its characterization over successive datasets could therefore be highly dependent on its sky location. In this study, we have observed two influences on binary characterization -- pulsar terms and PTA geometry. If the source's sky location is decently far from nearby pulsars, and the pulsars have reasonable distance precision, the source achieves signal detection and parameter estimation at low S/N, due to the significant separation in Earth and pulsar terms. If the source's sky location is close to many pulsars, it receives less pulsar term contribution at low S/N. However, as more pulsars with poorly-measured distances are added to the array (when the S/N is already at a moderate level), the PTA geometry becomes more influential. High measurement precision is then achieved at high S/N, where the pulsars' proximity to this source assists the parameter estimation. 

We note that this disparity between sky locations is more relevant for smaller arrays, for which the pulsar distribution is highly anisotropic. For larger PTAs, which tend to cover more of the sky, sources located near the bulk of pulsars will likely be best-characterized for all S/N. Given the competing influences on binary characterization, though, we find that the S/N is often an insufficient statistic for describing and predicting a source’s parameter estimates over time. Finally, since our simulated array generally reflects the current state of pulsar distance measurements, we can expect that CW characterization in real PTA datasets may be more dependent on array geometry than pulsar term effects.

The intention of this study was to investigate the evolution of a realistic CW detection; for example, we continually added pulsars to the array, gradually increased the dataset timespan, used up-to-date pulsar distance measurements, and injected real intrinsic pulsar red noise characteristics. While outside the scope of this work, further studies with simpler PTA configurations are needed in order to isolate the competing influences on CW parameter estimation, and localization in particular. Multi-messenger detections necessarily hinge on GW localization areas; the better localized a source, the better our chances of SMBHB host galaxy identification. Determining the relative importance of the contributing factors, and how their contributions may evolve over the accumulation of PTA data, will better prepare us for the single-source detection and characterization era.

\begin{acknowledgments}
PP and LS would like to thank Kyle Gersbach, William Lamb, Celia Fielding, and Matt Miles for their ideas and helpful discussions. We also thank Bence Bécsy for useful feedback on the manuscript. PP acknowledges support from NASA FINESST grant number 80NSSC23K1442. LS is greatly appreciative of support from NSF AST2307719 and NRT-2125764. MC acknowledges support by the European Union (ERC, MMMonsters, 101117624). The authors are members of the NANOGrav collaboration, which receives support from NSF Physics Frontiers Center award number 1430284 and 2020265. This work was conducted in part using the resources of the Advanced Computing Center for Research and Education (ACCRE) at Vanderbilt University, Nashville, TN. This work utilized the software suites \texttt{astropy} \cite{astropy:2013, astropy:2018, astropy:2022}, \texttt{HEALPix} \cite{2005ApJ...622..759G}, \texttt{healpy} \cite{Zonca2019}, \texttt{Jupyter} \cite{soton403913}, \texttt{La Forge} \cite{2020zndo...4152550S}, \texttt{Matplotlib} \cite{Hunter:2007},  \texttt{Numpy} \cite{harris2020array}, and \texttt{Scipy} \cite{2020SciPy-NMeth}.
\end{acknowledgments}

\appendix

\section{\label{sec:app}Additional figures}

\begin{figure*}[!ht]
     \centering
     \includegraphics[width=0.98\textwidth]{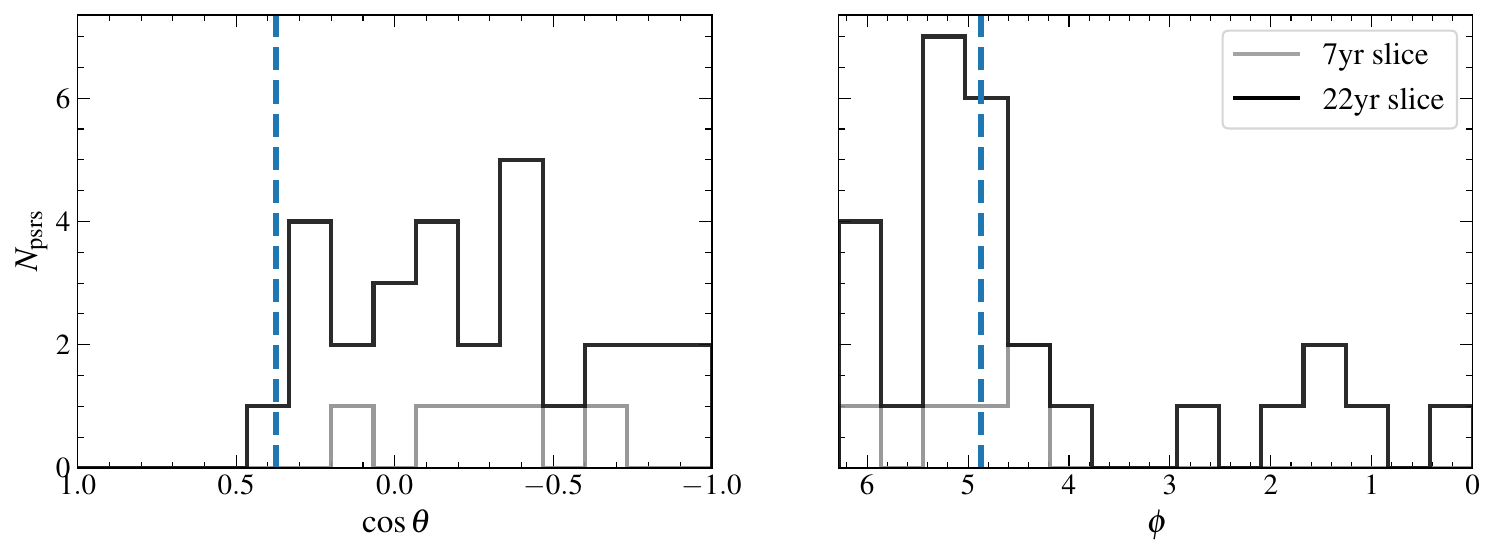}
     \caption{Histograms of pulsar sky locations in $\cos\theta$ (left panel) and $\phi$ (right panel) for early and late time slices. The $\cos\theta$ distributions are shown only for pulsars lying in a slice of R.A. centered on Location A, $\alpha = 279^{\circ} \pm 15^{\circ}$. Likewise, the $\phi$ distributions are shown only for pulsars lying in a slice of Dec. centered on Location A, $\delta = +22^{\circ} \pm 15^{\circ}$. The 7-year slice (20 pulsars) is shown in light gray, and the 22-year slice (all 116 pulsars) is shown in dark gray. Location A is indicated by the dashed blue line. Note that the parameter ranges decrease from left to right for easier comparison with the sky map in \autoref{fig:skymap}; the top (bottom) of the sky map corresponds to $\cos\theta=1$ ($\cos\theta=-1$), and the left-hand (right-hand) side of the sky map corresponds to $\phi=2\pi$ ($\phi=0$).}
     \label{fig:Acov}
\end{figure*}

\begin{figure*}[!ht]
     \centering
     \includegraphics[width=0.98\textwidth]{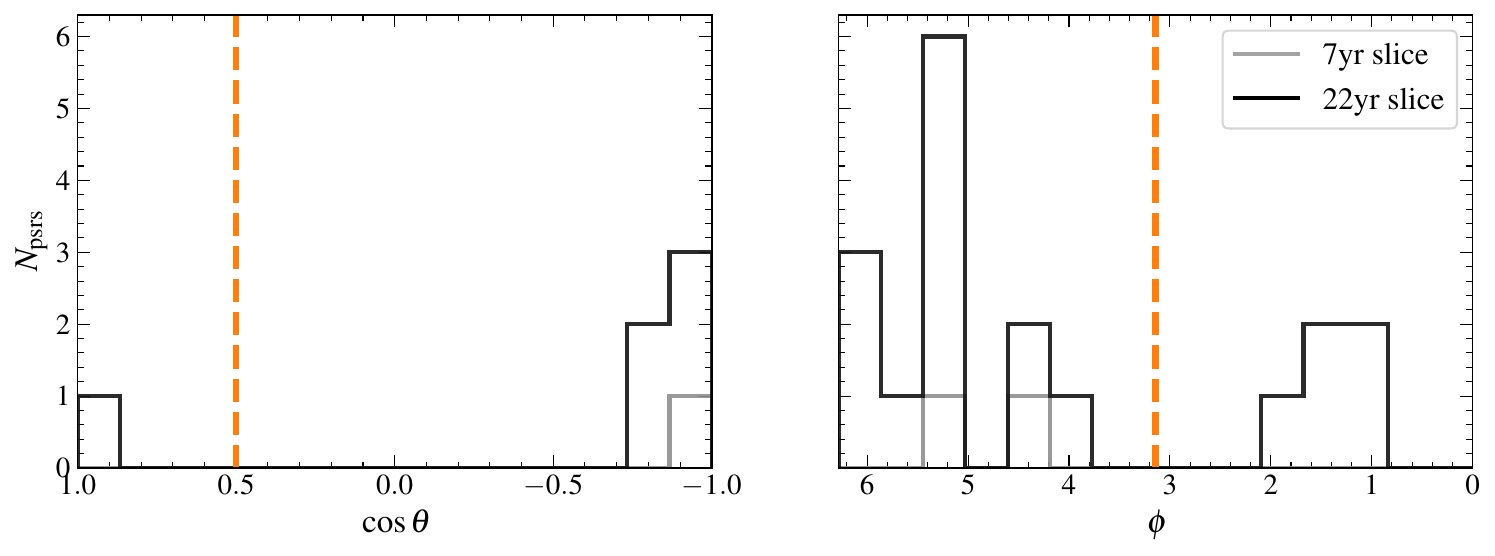}
     \caption{The same histograms as in \autoref{fig:Acov}, but now showing pulsar sky locations as they pertain to Location B. The $\cos\theta$ distributions are shown only for pulsars lying in a slice of R.A. centered on Location B, $\alpha = 180^{\circ} \pm 15^{\circ}$, and the $\phi$ distributions are shown only for pulsars lying in a slice of Dec. centered on Location B, $\delta = +30^{\circ} \pm 15^{\circ}$. Location B is indicated by the dashed orange line.}
     \label{fig:Bcov}
\end{figure*}

\begin{figure*}[!ht]
     \centering
     \includegraphics[width=0.98\textwidth]{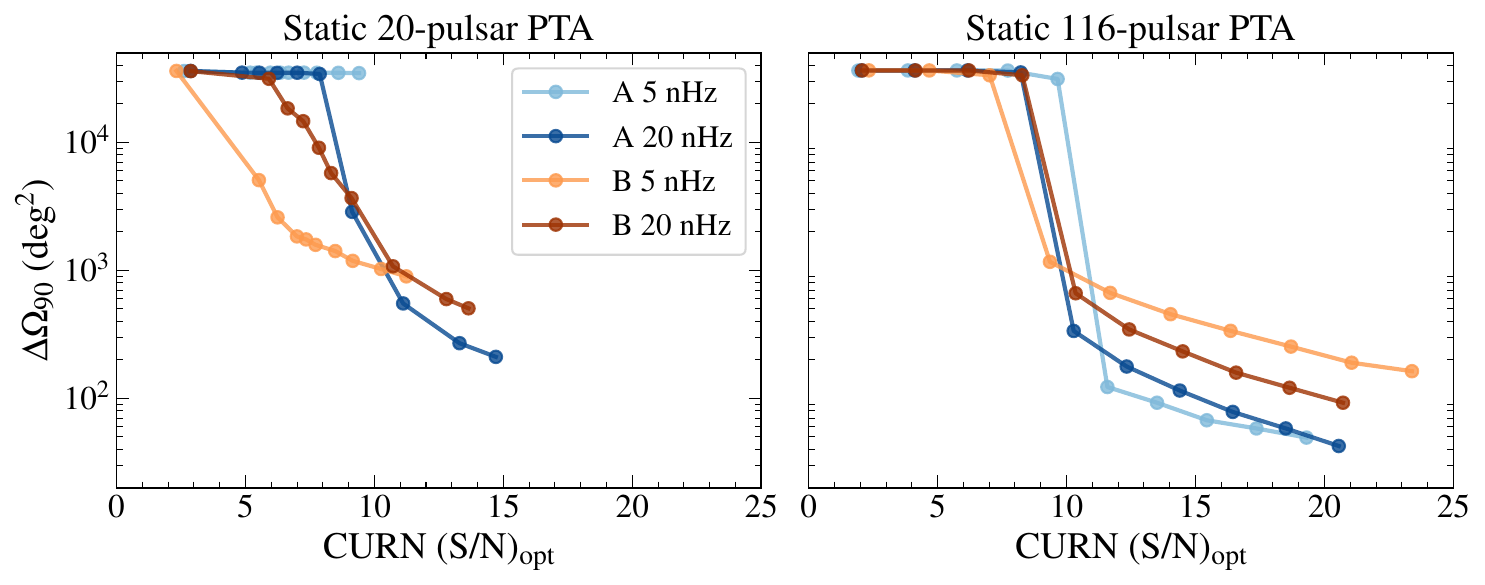}
     \caption{90\% credible area results from the two additional tests described in \autoref{sec:othertests}. As in \autoref{fig:fs_vs_eto}, the injections in these tests are noiseless, but the noise parameters are fixed in the PTA covariance matrix during sampling. Sources B5 and B20 are constrained at lower (S/N)$_{\rm opt}$ in the 20-pulsar array than in the 116-pulsar array, owing to the larger fraction of well-measured pulsar distances, and thus more contribution from pulsar terms. On the other hand, all sources achieve better localization precision at high (S/N)$_{\rm opt}$ in the 116-pulsar array, owing to pulsar geometry effects. Side note: source A5 in the static 20-pulsar PTA is not constrained in any time slice, reaching a maximum of only (S/N)$_{\rm opt} \sim 9$.}
     \label{fig:staticPTA_tests}
\end{figure*}

\begin{figure}[!ht]
     \centering
     \includegraphics[width=0.48\textwidth]{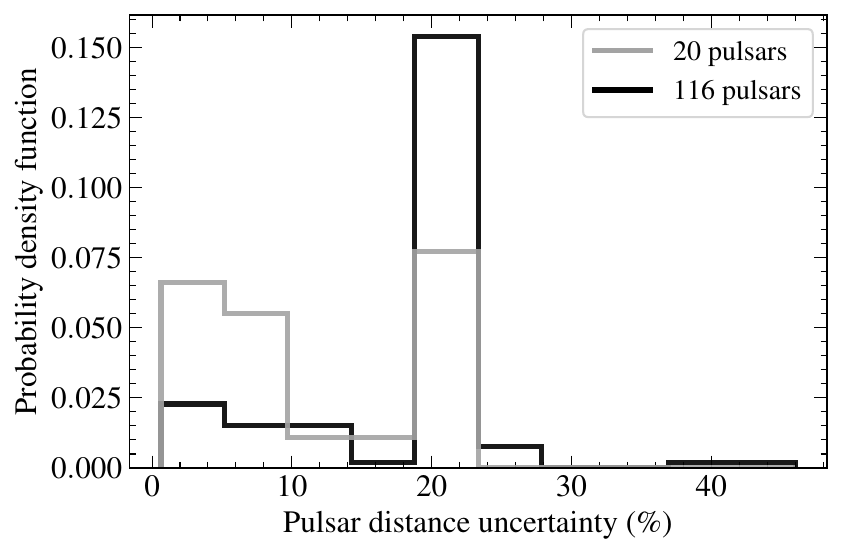}
     \caption{Distributions of pulsar distance uncertainties for the 20-pulsar and 116-pulsar arrays. The 20-pulsar array (corresponding to the 7-year slice) is shown in light gray, and the 116-pulsar array (corresponding to the 20-year and 22-year slices) is shown in dark gray.}
     \label{fig:pdist_uncs}
\end{figure}

\bibliography{delphi2}

\end{document}